\documentclass[prb,amsmath,twocolumn]{revtex4-2}

\usepackage{bm}
\usepackage{dsfont}
\usepackage{physics}
\usepackage{mathtools}
\usepackage{csquotes}
\usepackage{multirow}
\usepackage{color}

\usepackage{hyperref}
\hypersetup{
	colorlinks=true,
	citecolor=blue, 
	linkcolor=blue, 
	urlcolor=blue,  
}

\newcommand{\bk}{{\bm k}}
\renewcommand{\eqref}[1]{Eq.~(\ref{#1})}

\begin{document}
\title{Origin of the anomalous Hall effect in two-band chiral
  superconductors}
\date{\today}

\author{M. D. E. Denys}
\email{mathew.denys@postgrad.otago.ac.nz}

\author{P. M. R. Brydon}
\email{philip.brydon@otago.ac.nz}

\affiliation{Department of Physics and MacDiarmid Institute for Advanced Materials and Nanotechnology, University of Otago, PO Box 56, Dunedin 9054, New Zealand}
	
\begin{abstract}
We consider the origin of the anomalous Hall effect in a general model of a clean two-band chiral superconductor. Within the Kubo formalism we derive an analytic expression for the high-frequency ac Hall conductivity valid close to the critical temperature. This expression involves two distinct gauge-invariant time-reversal-odd bilinear (TROB) functions involving the pairing potential and its Hermitian conjugate. We argue that the existence of at least one of these TROBs generically implies a nonzero ac Hall conductivity. The TROBs allow us to clarify the roles of intra- and interband pairing, and provide a straightforward criterion for a superconducting state to exhibit the anomalous Hall effect. We briefly exemplify our results with model calculations for a chiral $p$-wave pairing state in strontium ruthenate and a chiral $d$-wave pairing state on the honeycomb lattice.
\end{abstract}

\maketitle

\section{Introduction}
Chiral superconductivity is an exotic pairing state characterized by the spontaneous breaking of time-reversal symmetry and a handed winding of the gap phase around the Fermi surface~\cite{kallin_chiral_2016}. Chiral pairing states have been proposed for a number of superconductors, most notably Sr$_2$RuO$_4$ \cite{mackenzie_superconductivity_2003}, UPt$_3$ \cite{norman_1992,joynt_2002}, URu$_2$Si$_2$ \cite{kasahara_exotic_2007}, and twisted bilayer graphene \cite{cao_unconventional_2018}, although a definitive interpretation of the experimental evidence remains elusive. The observation of the polar Kerr effect, which is closely related to the anomalous Hall effect (AHE) \cite{argyres_theory_1955,kapitulnik_polar_2009}, is a key signature of time-reversal symmetry breaking (TRSB) in the bulk. Indeed, a nonzero Kerr signal has been observed in various candidate superconductors, such as Sr$_2$RuO$_4$ \cite{xia_high_2006}, UPt$_3$ \cite{schemm_observation_2014}, URu$_2$Si$_2$ \cite{schemm_evidence_2015}, Bi/Ni bilayers \cite{gong_time-reversal_2017}, PrO$_4$Sb$_{12}$ \cite{LevensonFalk_2018}, and UTe$_2$~\cite{Hayes_UTe2}, providing strong evidence for the presence of chiral pairing states in these materials. 

The origin of the polar Kerr effect in a chiral superconductor has been the subject of much debate, as the breaking of time-reversal symmetry by the pairing potential is not sufficient to explain the presence of the AHE. Specifically, in chiral superconductors time-reversal symmetry is broken in the relative momentum coordinate of the Cooper pair, while it is the center-of-mass coordinate which couples to the external electric field in the AHE \cite{kallin_chiral_2016,taylor_intrinsic_2012}. Thus, the AHE is vanishing in single-band superconductors except in the presence of impurities that break the translational symmetry which necessitates that these coordinates are independent \cite{konig_kerr_2017,goryo_impurity-induced_2008,lutchyn_frequency_2009,li_anomalous_2015,Zhang_AHE_2019}. On the other hand, the relative and center-of-mass coordinates are coupled in multiband superconductors \cite{taylor_intrinsic_2012}, so mechanisms intrinsic to the clean superconductor can contribute to the AHE. Such an intrinsic contribution has been theoretically demonstrated in a large number of models \cite{wysokinski_intrinsic_2012,taylor_intrinsic_2012,gradhand_kerr_2013,wang_intrinsic_2017,robbins_effect_2017,joynt_superconductivity_2017,Triola_2018,brydon_loop_2019,Zhang_SRO_2020}. Nevertheless, it is unclear what conditions the pairing potential in a multiband model should satisfy for the existence of the AHE.

Within a simple model of chiral $d$-wave pairing on the honeycomb lattice, Ref.~\cite{brydon_loop_2019} identified a TRSB bilinear combination of the pairing potential and its Hermitian conjugate as critical to the appearance of the AHE. Dubbed the ``time-reversal-odd bilinear'' (TROB), this quantity is explicitly gauge-invariant and breaks time-reversal symmetry by construction. Although the simplicity of the model studied in Ref.~\cite{brydon_loop_2019} made the wider relevance of the TROB uncertain, the authors speculated that a nonzero TROB is crucial for the existence of an intrinsic AHE in a multiband chiral superconductor. In this manuscript we demonstrate that this is indeed the case by analytically calculating the ac Hall conductivity for a generic model of a two-band superconductor. Our work establishes general conditions on the form of the pairing potential required for the existence of a TROB.

We begin in Sec.~\ref{sec:model} by introducing a general model of a two-band system where the normal state Hamiltonian includes all terms allowed by inversion and time-reversal symmetry. Using the Kubo formalism, in Sec.~\ref{sec:TROBs} we analytically determine the leading contribution to the ac Hall conductivity in the limit of high frequencies and a small gap, revealing that the Hall conductivity depends upon two distinct TROBs. We evaluate these TROBs for arbitrary even- and odd-parity pairing states in Sec.~\ref{subsec:genericTROB}, which reveals the general importance of interband pairing, but suggests that purely-intraband pairing may also play a role. Although our analysis is performed in an asymptotic limit, we argue in Sec.~\ref{sec:generalizations} that our conclusions should hold more generally. Our analysis is made concrete in Sec.~\ref{sec:examples} by considering two model systems: chiral $p$-wave pairing in Sr$_2$RuO$_4$~\cite{taylor_intrinsic_2012} and $d$-wave superconductivity on the honeycomb lattice~\cite{brydon_loop_2019}. Finally, we conclude in Sec.~\ref{sec:conclusions} with an outlook for future work.

\section{Model} \label{sec:model}
We consider a general superconducting system in which the electronic states are described by their momentum, spin, and an additional discrete degree of freedom which we denote as the ``orbital''. In a model with two orbitals the single-particle Hamiltonian is written in the Bogoliubov de Gennes (BdG) formalism as
\begin{equation} \label{eq:ham_spinorform}
\mathcal H = \frac{1}{2}\sum_{\bk} \Psi_\bk^\dagger
	\begin{pmatrix}
		H_0(\bk)            & \Delta(\bk)  \\
		\Delta^\dagger(\bk) & -H_0^T(-\bk)
	\end{pmatrix}
	\Psi_\bk,
\end{equation}
where 
\begin{equation}
\Psi_\bk = \begin{pmatrix} \hat{\bf a}_{\bk} \\ \hat{\bf a}^\ast_{-{\bf k}} \end{pmatrix}
\end{equation}
with 
\begin{equation}
\hat{\bf a}_{\bf k}^T =
	(\hat a_{\bk,1,\uparrow},
	 \hat a_{\bk,1,\downarrow},
	 \hat a_{\bk,2,\uparrow},
	 \hat a_{\bk,2,\downarrow}),
\end{equation}
and $\hat a_{\bk,\eta,\sigma}$ the annihilation operator for an electron with momentum $\bk$ and spin $\sigma$ in the orbital $\eta$. Both $H_0$ and $\Delta$ are $4\times4$ matrices in the combined orbital-spin space, and transform under inversion ${\cal I}$ and time-reversal $\Theta$ as
\begin{align}
\mathcal I: H_0(\bk)    & = U_I^\dagger H_0(-\bk) U_I = H_0(\bk), \label{eq:H0_I}                 \\
\mathcal I: \Delta(\bk) & = U_I^\dagger \Delta(-\bk) U_I^* = \pm \Delta(\bk), \label{eq:Delta_I}  \\
\Theta: H_0(\bk)        & = U_T^\dagger H_0^*(-\bk) U_T = H_0(\bk), \label{eq:H0_T}               \\
\Theta: \Delta(\bk)     & = U_T^\dagger \Delta^*(-\bk) U_T^* \neq \Delta(\bk), \label{eq:Delta_T}
\end{align}
where $U_I$ and $U_T$ are unitary matrices. Whereas the normal state Hamiltonian $H_0$ is assumed to be symmetric under both spatial inversion and time-reversal, the pairing potential $\Delta(\bk)$ is assumed to have definite parity but breaks time-reversal symmetry. Although the derivation in Section~\ref{sec:TROBs} proceeds identically for mixed-parity pairing states, we ignore this case here since it typically requires broken inversion symmetry in the normal state.

The normal state Hamiltonian matrix can be decomposed as
\begin{equation} \label{eq:H0Paulidecomosition}
H_0(\bk) = \sum_{\alpha,\beta=0}^3 h_{\alpha\beta}(\bk) \ \eta_\alpha \otimes \sigma_\beta,
\end{equation}
where $\eta_\alpha$ ($\sigma_\beta$) are Pauli matrices encoding the orbital (spin) degree of freedom. Hermiticity requires the $h_{\alpha\beta}(\bk)$ coefficients to be real-valued, while the presence of inversion and time-reversal permit just six $(\alpha,\beta)$ pairs in \eqref{eq:H0Paulidecomosition}. These correspond to five mutually anticommuting matrices $\gamma_{1\ldots5}$, along with the identity matrix $\mathds 1_4=\eta_0\otimes\sigma_0$.  The particular form of the matrices $\gamma_i$ depends on the system under consideration. For example, a system with two orbitals of the same parity has $U_I=\mathds 1_4$ and $U_T=\eta_0\otimes i\sigma_2$, and the allowed terms in \eqref{eq:H0Paulidecomosition} are
\begin{equation} \label{alpha_beta_terms_same_partiy}
(\alpha,\beta) = (0,0),\,\, (1,0),\,\, (3,0),\,\, (2,1),\,\, (2,2),\,\,  (2,3).
\end{equation} 
For a system with orbitals exchanged by inversion symmetry such that $U_I=\eta_1\otimes\sigma_0$ and $U_T=\eta_0\otimes i\sigma_2$, the allowed terms are 
\begin{equation}
(\alpha,\beta) = (0,0), \,\, (2,0), \,\, (3,0), \,\, (1,1), \,\, (1,2), \,\, (1,3).
\end{equation} All other possibilities are summarized in Table \ref{tab:alpha_beta_vals}. Although some of these may appear atypical, they can each be related to one of the cases noted above via a canonical transformation.

\begin{table}
	\centering
	\begin{tabular}{|c|c||c|}
		\hline
		   $U_I^{[\mathrm o]}$    & $U_T^{[\mathrm o]}$ &      Allowed $(\alpha,\beta)$ terms      \\ \hline
		\multirow{3}{*}{$\eta_0$} &      $\eta_0$       & (0,0), (1,0), (3,0), (2,1), (2,2), (2,3) \\
		                          &      $\eta_1$       & (0,0), (1,0), (2,0), (3,1), (3,2), (3,3) \\
		                          &      $\eta_3$       & (0,0), (2,0), (3,0), (1,1), (1,2), (1,3) \\ \hline
		\multirow{2}{*}{$\eta_1$} &      $\eta_0$       & (0,0), (1,0), (2,0), (3,1), (3,2), (3,3) \\
		                          &      $\eta_1$       & (0,0), (1,0), (3,0), (2,1), (2,2), (2,3) \\ \hline
		\multirow{2}{*}{$\eta_2$} &      $\eta_1$       & (0,0), (2,0), (3,0), (1,1), (1,2), (1,3) \\
		                          &      $\eta_3$       & (0,0), (1,0), (2,0), (3,1), (3,2), (3,3) \\ \hline
		\multirow{2}{*}{$\eta_3$} &      $\eta_0$       & (0,0), (2,0), (3,0), (1,1), (1,2), (1,3) \\
		                          &      $\eta_3$       & (0,0), (1,0), (3,0), (2,1), (2,2), (2,3) \\ \hline
	\end{tabular}
	\caption{The $(\alpha,\beta)$ terms permitted in \eqref{eq:H0Paulidecomosition} for systems with each possible form of inversion and time-reversal,  enumerated by the orbital parts of $U_I$ and $U_T$. In each case the allowed terms correspond to the identity along with five mutually anticommuting matrices. $U_T^{[\mathrm o]}=\eta_2$ is not allowed as it leads to $U_TU_T^*=+\mathds 1$, which does not correspond to a spin-half system. We exclude cases where inversion does not commute with time-reversal (i.e. $U_IU_T\neq U_TU_I^*$) and where the inversion operator is not its own inverse (i.e. $U_IU_I\neq\mathds 1$).}
	\label{tab:alpha_beta_vals}
\end{table}

It is convenient to re-write the normal state Hamiltonian as
\begin{equation} \label{eq:H0gammadecomosition}
H_0(\bk) = h_{00}(\bk) \mathds 1_4 + \bm h(\bk) \cdot \bm \gamma,
\end{equation}
where $h_{00}(\bk)$ is an even function of momentum, $\bm \gamma~=~ (\gamma_1,\ldots,\gamma_5)$ is a vector of the gamma matrices, and $\bm h(\bk)=(h_{\alpha_1\beta_1},\ldots,h_{\alpha_5\beta_5})$ is a vector of the corresponding coefficients. For future reference we define the \enquote{flattened} normal state Hamiltonian
\begin{equation} \label{eq:H0flattened}
\tilde H_0(\bk) = \frac{H_0(\bk) - h_{00}(\bk)\mathds 1_4}{\abs{\bm h(\bk)}} = \hat{\bm h}(\bk) \cdot \bm \gamma,
\end{equation}
where $\hat{\bm h} = \bm h / \abs{\bm h}$. This is traceless and satisfies the eponymous property
\begin{equation} \label{eq:H0tildeeigenequation}
\tilde H_0(\bk) \ket{\bk,\pm,s} = \pm\ket{\bk,\pm,s},
\end{equation}
where $\ket{\bk,\pm,s}$ are eigenstates of $H_0$ with the good quantum numbers momentum $\bk$, band $\pm$, and pseudospin $s$. The existence of a degenerate pseudospin index is guaranteed by the inversion and time-reversal symmetries \cite{yip_pseudospin_2016,venderbos_odd-parity_2016}

Analogously to \eqref{eq:H0Paulidecomosition}, the pairing potential $\Delta$ can be decomposed in terms of orbital and spin Pauli matrices, but its form is not so restricted because inversion and time-reversal symmetries can be broken in the superconducting state. We only enforce that the potential satisfies fermionic antisymmetry, requiring that $\Delta(\bk) = - \Delta^T(-\bk)$. For convenience, we define the transformed pairing potential
\begin{equation} \label{eq:deltatilde}
\tilde \Delta (\bk) = \Delta(\bk) U_T^\dagger,
\end{equation}
which transforms analogously to $H_0$ under point symmetry operations, and has the useful property that $\tilde \Delta$ and $\tilde \Delta^\dagger$ are time-reversed counterparts. Although the particular form of $\tilde\Delta(\bk)$ is set by the details of the system, we note that for an even-parity pairing state it will be a linear combination of the six $\eta_\alpha\otimes\sigma_\beta$ matrices which are allowed to appear in the normal state Hamiltonian; the potential for an odd-parity pairing state will involve the other ten matrices.

\section{Hall conductivity and Time-reversal-odd bilinears} \label{sec:TROBs}
The frequency-dependent Hall conductivity is defined as 
\begin{equation} \label{eq:hall_def}
\sigma_H(\omega) = \frac{i}{2\omega} \lim_{i\omega_n \to \omega+i0^+} [ \pi_{xy}(i\omega_n) - \pi_{yx}(i\omega_n)],
\end{equation}
where $\pi_{ab}$ is the current-current correlation function \cite{mahan_many-particle_1980}
\begin{equation}
\pi_{ab}(i\omega_n) = -\frac{1}{N}\int^\beta_0 d\tau \, e^{i\omega_n\tau}\langle T_\tau J_a(\tau)J_b(0)\rangle,
\end{equation}
$N$ is the number of lattice points in the crystal, $\tau$ is the imaginary time, $J_a$ is the $a^{\text{th}}$ component of the current operator  
\begin{equation}
J_a = e\sum_{{\bk}}\Psi^\dagger_{\bk}v^{a}_{\bk}\Psi_{\bk},
\end{equation}
and the velocity matrix $v^{a}_{\bk}$ is given by
\begin{equation} \label{eq:velocityoperator}
v^a_\bk =
	\begin{pmatrix}
		v^{ea}_\bk & 0 \\
		0 & v^{ha}_\bk
	\end{pmatrix}
  = \begin{pmatrix}
	  \frac{\partial H_0(\bk)}{\partial k_a} & 0 \\
	  0 & \frac{\partial (-H^T_0(-\bk))}{\partial k_a}
	\end{pmatrix}.
\end{equation}
In the linear response regime $\pi_{ab}(i\omega_n)$ is evaluated at the one-loop level, yielding
\begin{multline} \label{eq:correlator_exact}
\pi_{ab}(i\omega_n) = \frac{e^2}{2N\beta} \sum_{\bk,m} \Tr{v_\bk^a \mathcal G_{\bk,i\omega_n+i\nu_m}v^b_\bk \mathcal G_{\bk,i\nu_m}},
\end{multline}
where $\mathcal G_{\bk,i\nu_m}$ is the Matsubara Green's function corresponding to \eqref{eq:ham_spinorform}. Note that in \eqref{eq:correlator_exact} $\omega_n=2n\pi/\beta$ is a Bose Matsubara frequency, and $\nu_m=(2m+1)\pi/\beta$ is a Fermi Matsubara frequency. For convenience we set $\hbar=1$. 

While \eqref{eq:correlator_exact} can be evaluated for an arbitrary pairing potential, the resulting analytic expression is generally very complicated and offers only limited insight. Progress can be made by considering the high-frequency, small-gap limit, obtained by neglecting terms in \eqref{eq:correlator_exact} of lower order than $\omega^{-2}$ and higher order than $\abs{\Delta}^2$. This choice is justified since experiments are often performed in the high-frequency regime, and the gap is small compared to other relevant energy scales; the small-gap approximation is expected to be particularly accurate close to the critical temperature. Performing a diagrammatic expansion of~\eqref{eq:correlator_exact}, an approximate expression for the intrinsic anomalous Hall conductivity in this limit is given by \cite{brydon_loop_2019}
\begin{multline} \label{eq:hall_smallgap}
\sigma_H(\omega) \approx \frac{ie^2}{2N\omega^2\beta} \sum_{\bk,m} \Tr\{[v \wedge v] \mathcal G_0 H_\Delta \mathcal G_0 H_\Delta \mathcal G_0\},
\end{multline}
where $[a \wedge b] = a^xb^y-a^yb^x$, and explicit momentum and frequency indices have been dropped for convenience. $H_\Delta$ is the pairing part of the BdG Hamiltonian,
\begin{equation}
H_{\Delta,\bk} =
	\begin{pmatrix}
		0 & \Delta(\bk) \\ \Delta^\dagger(\bk) & 0
	\end{pmatrix},
\end{equation}
and $\mathcal G_0 = \mathcal G_{0,\bk,i\nu_m}$ is the normal state Matsubara Green's function in the Nambu representation
\begin{equation}
\mathcal G_{0,\bk,i\nu_m} =
	\begin{pmatrix}
		\mathcal G^e_{0,\bk,i\nu_m} & 0 \\
		0 & \mathcal G^h_{0,\bk,i\omega_n}
	\end{pmatrix}.
\end{equation}
With these expressions, the trace in \eqref{eq:hall_smallgap} is split into two terms. By enforcing that $H_0$ is symmetric under both inversion and time-reversal, and noting that time-reversal commutes with inversion (i.e $U_I U_T = U_T U_I^*$, as time-reversal is antiunitary), it can be shown that $[v^h \wedge v^h] = - U_I^T [v^e \wedge v^e]^T U_I^*$ and $U_I^\dagger U_T^\dagger [v^e \wedge v^e]^T U_T U_I = -[v^e \wedge v^e]$, from which we obtain
\begin{widetext}
\begin{align}
\sigma_H(\omega) \approx \frac{ie^2}{2N\omega^2\beta} \sum_{\bk,m} \left[\Tr\big\{[v^e \wedge v^e] \mathcal G_0^e \tilde\Delta U_T \mathcal G_0^h U_T^\dagger \tilde\Delta^\dagger \mathcal G_0^e\big\} + \Tr\big\{[v^e \wedge v^e] U_T^\dagger \mathcal G_0^h U_T \tilde\Delta^\dagger \mathcal G_0^e \tilde\Delta U_T^\dagger \mathcal G_0^h U_T \big\}\right] \label{eq:hall4}
\end{align}
\end{widetext}
in the high-frequency, small-gap limit. The electron-like Green's functions can be written in terms of projection operators onto each normal state energy band:
\begin{equation}
\mathcal G^e_{0,\bk,i\nu_m} = \sum_{\pm} \frac{\mathcal P_{\bk,\pm}}{i\nu_m-E_{\bk,\pm}}.
\end{equation}
The projection operators are defined in terms of  \eqref{eq:H0flattened} as $\mathcal P_{\bk,\pm} = (\mathds 1 \pm \tilde H_0(\bk))/2$, and therefore
\begin{align}
\mathcal G^e_{0,\bk,i\nu_m}
&= \frac{\mathds 1 + \tilde H_0(\bk)}{2(i\nu_m-E_{\bk,+})} +
  \frac{\mathds 1 - \tilde H_0(\bk)}{2(i\nu_m-E_{\bk,-})} \notag \\
&= a_{\bk,\nu_m} \mathds 1 + b_{\bk,\nu_m} \tilde H_0(\bk), \label{eq:Ge}
\end{align}
where
\begin{align}
a_{\bk,\nu_m} &= \frac{1}{2} \left[ (i\nu_m - E_{\bk,+})^{-1} + (i\nu_m - E_{\bk,-})^{-1} \right], \label{eq:a_def}\\
b_{\bk,\nu_m} &= \frac{1}{2} \left[ (i\nu_m - E_{\bk,+})^{-1} - (i\nu_m - E_{\bk,-})^{-1} \right].
\end{align}
Analogously, the hole-like Green's function has the form
\begin{equation}
\mathcal G^h_{0,\bk,i\nu_m} = c_{\bk,\nu_m} \mathds 1 + d_{\bk,\nu_m} \tilde H_0^T(\bk), \label{eq:Gh}
\end{equation}
where $c_{\bk,\nu_m}=-a_{\bk,-\nu_m}$ and $d_{\bk,\nu_m}=-b_{\bk,-\nu_m}$. Further, because $H_0$ is symmetric under time-reversal,
\begin{equation}
U_T^\dagger \mathcal G^h_{0,\bk,i\nu_m} U_T = c_{\bk,\nu_m} \mathds 1 + d_{\bk,\nu_m} \tilde H_0(\bk). \label{eq:Gh_TR}
\end{equation}
Substituting \eqref{eq:Ge} and \eqref{eq:Gh_TR} into \eqref{eq:hall4}, and using the relationship between $a_{\bk,\nu_m}$ ($b_{\bk,\nu_m}$) and $c_{\bk,\nu_m}$ ($d_{\bk,\nu_m}$), we obtain after some manipulation
\begin{widetext}
\begin{multline} \label{eq:TROBterm}
\sigma_H(\omega) = \frac{ie^2}{2N\omega^2\beta} \sum_{\bk,m} \Big[
	 a^2c \Tr \{ [v^e\! \wedge v^e] \ \mathrm{TROB}_1\}
	+b^2c \Tr \{ \tilde{H}_0 [v^e\! \wedge v^e] \tilde{H}_0 \ \mathrm{TROB}_1\}
	+abc  \Tr \{ \{\tilde{H}_0, [v^e\! \wedge v^e]\} \ \mathrm{TROB}_1\} \\
	+a^2d \Tr \{ [v^e\! \wedge v^e] \ \mathrm{TROB}_2\}
	+b^2d \Tr \{ \tilde{H}_0 [v^e\! \wedge v^e] \tilde{H}_0 \ \mathrm{TROB}_2\}
	+abd  \Tr \{ \{\tilde{H}_0, [v^e\! \wedge v^e]\} \ \mathrm{TROB}_2\} \Big],
\end{multline}
\end{widetext}
where momentum and frequency indices are suppressed for clarity. The key feature of \eqref{eq:TROBterm} is the introduction of two distinct time-reversal-odd bilinears (TROBs), which are defined in terms of the pairing potential:
\begin{align}
\mathrm{TROB}_1 &\equiv \tilde\Delta \tilde\Delta^\dagger - \tilde \Delta^\dagger \tilde \Delta, \label{eq:TROB1}\\
\mathrm{TROB}_2 &\equiv \tilde\Delta \tilde H_0 \tilde\Delta^\dagger - \tilde \Delta^\dagger \tilde H_0 \tilde \Delta. \label{eq:TROB2}
\end{align}
TROB$_1$ was initially introduced in \cite{brydon_loop_2019}. It can be verified that both TROBs are the difference of a gauge-invariant quantity and its time-reversed value, making them odd under time-reversal by construction. The significance of the TROBs is that they translate the TRSB in the particle-particle channel to the observable particle-hole channel. 

This direct dependence of the anomalous Hall conductivity on TROBs is the core result of this paper. We observe that the presence of at least one non-vanishing TROB is a necessary condition for a non-zero Hall conductivity in the high-frequency, small-gap limit. Although this is not a \emph{sufficient} condition, fine-tuning is required for
\eqref{eq:TROBterm} to be vanishing if either TROB is non-zero. Importantly, although the TROBs are only nonzero for a TRSB  pairing potential, a vanishing TROB does not imply that the pairing potential is time-reversal symmetric. As such, the TROBs significantly constrain which superconducting states can give rise to a non-zero anomalous Hall conductivity. Both TROBs are straightforward to calculate, and hence provide an easy mechanism for identifying candidate Hamiltonians worth further examination.

Another important aspect of~\eqref{eq:TROBterm} is the presence of the wedge product of velocity matrices $[v^e \wedge v^e]$. Since this product is vanishing in a single-band system, it implies that a multiband model is necessary for an anomalous Hall conductivity.

\subsection{Interband vs intraband pairing} \label{subsec:genericTROB}
Interband pairing has been identified by a number of authors as another necessary condition for an AHE \cite{taylor_intrinsic_2012,brydon_loop_2019}. This is notable because intraband pairing generally guarantees the robustness of the superconducting instability, while intraband pairing competes, and is therefore detrimental to the formation of a superconducting state \cite{ramires_tailoring_2018}. The connection between, and compatibility of, these claims and our result is of interest. Here we show that the TROBs directly evidence the key role played by interband pairing, while also demonstrating that purely intraband pairing can generate a nonzero Hall conductivity.

\subsubsection{Even-parity pairing}
The general form of an even-parity pairing potential in the band-pseudospin basis is
\begin{align} \label{eq:Deltae}
\tilde{\Delta}_e &          =    \frac{1}{2}(\psi_++\psi_-)b_0\otimes s_0 + \frac{1}{2}(\psi_+-\psi_-)b_3\otimes s_0  \notag \\
	             & \phantom{=} + \psi_eb_1\otimes s_0 + {\bf d}_e\cdot(b_2\otimes {\bf s}),
\end{align}
where the first line describes intraband pseudospin-singlet pairing with potential $\psi_\pm$ in band $\pm$, while $\psi_e$ and ${\bf d}_e$ are the interband pseudospin-singlet and -triplet potentials. The Pauli matrices $b_\mu$ and $s_\mu$ encode the band and pseudospin degrees of freedom, respectively, and all pairing potentials are even functions of the momentum. Because the pseudospin and band indices are even under inversion, the inversion operator is trivial in this representation, and so the matrices appearing in \eqref{eq:Deltae} are essentially the same as in the case of \eqref{alpha_beta_terms_same_partiy}. The two TROBs evaluate as 
\begin{align}
	\text{TROB}_1 & =  2i{\bf d}_e\times{\bf d}_e^\ast\cdot(b_0\otimes{\bf s}) - 4\text{Im}\{\psi_e{\bf d}_e^\ast\}\cdot(b_3\otimes{\bf s}) \notag \\
	              & \phantom{=} - 2\text{Im}\{(\psi_+-\psi_-)\psi_e^\ast\}b_2\otimes s_0 \notag                                                    \\
	              & \phantom{=} + 2\text{Im}\{(\psi_+-\psi_-){\bf d}_e^\ast\}\cdot(b_1\otimes{\bf s}),                                             \\
	\text{TROB}_2 & = -2i{\bf d}_e\times{\bf d}_e^\ast\cdot(b_3\otimes{\bf s}) + 4\text{Im}\{\psi_e{\bf d}_e^\ast\}\cdot(b_0\otimes{\bf s}) \notag \\
	              & \phantom{=}- 2\text{Im}\{(\psi_++\psi_-)\psi_e^\ast\}b_2\otimes\sigma_0 \notag                                                 \\
	              & \phantom{=}+2\text{Im}\{(\psi_++\psi_-){\bf d}_e^\ast\}\cdot(b_1\otimes{\bf s}). \label{eq:TROB2_even}
\end{align}
We observe that the pairing potential must have interband components for either TROB to be nonzero, which is generically the case if the pairing potential has a nontrivial matrix structure when expressed in the orbital-spin basis. Note that the velocity operators are not generally diagonal in the band-pseudospin picture, and so the off-diagonal components of the TROBs are still relevant to the existence of the Hall conductivity. We note that the diagonal blocks of $\text{TROB}_1$ also play a crucial role in the inflation of point or line nodes into Bogoliubov Fermi surfaces~\cite{brydon_bfs_2018}.

\subsubsection{Odd-parity pairing}
Adopting the same band-pseudospin basis as above, a general odd-parity pairing potential has the form
\begin{align}
 \tilde{\Delta}_o &         =    \frac{1}{2}({\bf d}_++{\bf d}_-)\cdot(b_0\otimes {\bf s}) + \frac{1}{2}({\bf d}_+-{\bf d}_-)\cdot(b_3\otimes {\bf s}) \notag \\
				  &\phantom{=} + \psi_ob_2\otimes s_0 + {\bf d}_o\cdot(b_1\otimes {\bf s}),
\end{align}
where in addition to the intraband triplet potentials ${\bf d}_\pm$ there are also interband singlet $\psi_o$ and triplet ${\bf d}_o$ potentials. Here all potentials are odd in momentum. Evaluating the TROBs we obtain
\begin{align}
	\text{TROB}_1 & =  i{\bf d}_+\times{\bf d}_+^\ast\cdot([b_0+b_3]\otimes{\bf s}) \notag                                                                      \\
	              & \phantom{=} + i{\bf d}_-\times{\bf d}_-^\ast\cdot([b_0-b_3]\otimes{\bf s})\notag                                                            \\
	              & \phantom{=} +2i{\bf d}_o\times{\bf d}^\ast_o\cdot(b_0\otimes {\bf s}) + 4\text{Im}\{\psi_o{\bf d}^\ast_o\}\cdot(b_3\otimes {\bf s})\notag   \\
	              & \phantom{=} + 2\text{Im}\{{\bf d}_o\cdot[{\bf d}_+-{\bf d}_-]^\ast\} b_2\otimes s_0\notag                                                   \\
	              & \phantom{=} -2\text{Im}\{{\bf d}_o\times[{\bf d}_++{\bf d}_-]^\ast\}\cdot(b_1\otimes {\bf s}), \notag                                       \\
	              & \phantom{=} - 2\text{Im}\{\psi_o[{\bf d}_+-{\bf d}_-]^\ast\}\cdot(b_1\otimes {\bf s})                                                       \\
	\text{TROB}_2 & = i{\bf d}_+\times{\bf d}_+^\ast\cdot([b_0+b_3]\otimes{\bf s}) \notag                                                                       \\
	              & \phantom{=} - i{\bf d}_-\times{\bf d}_-^\ast\cdot([b_0-b_3]\otimes{\bf s})\notag                                                            \\
	              & \phantom{=} +2i{\bf d}_o\times{\bf d}^\ast_o\cdot(b_3\otimes {\bf s}) - 4\text{Im}\{\psi_o{\bf d}^\ast_o\}\cdot(b_0\otimes {\bf s})\notag \\
	              & \phantom{=} + 2\text{Im}\{{\bf d}_o\cdot[{\bf d}_++{\bf d}_-]^\ast\} b_2\otimes s_0\notag                                                   \\
	              & \phantom{=} -2\text{Im}\{{\bf d}_o\times[{\bf d}_+-{\bf d}_-]^\ast\}\cdot(b_1\otimes {\bf s}) \notag                                        \\
	              & \phantom{=} - 2\text{Im}\{\psi_o[{\bf d}_++{\bf d}_-]^\ast\}\cdot(b_1\otimes {\bf s}).
\end{align}
In contrast to the even-parity case, we observe that a purely-intraband potential with ${\bf d}_\pm\times{\bf d}_\pm^\ast\neq0$ can give rise to a nonzero TROB. Although such nonunitary pairing states are generically realized in TRSB single-band systems, the presence of the wedge product of velocity matrices in~\eqref{eq:TROBterm} is nevertheless only nonzero in a multiband system, reiterating the necessity of multiple bands for an anomalous Hall conductivity.

\subsection{Away from the high-frequency, small-gap limit} \label{sec:generalizations}
Although~\eqref{eq:TROBterm} only  rigorously applies in the high-frequency, small-gap limit, the conclusion that a nonzero TROB implies a nonzero Hall conductivity is generically valid, since any correction terms are unlikely to exactly cancel the leading-order contribution~\eqref{eq:TROBterm}. In considering the converse   statement, we start by noting that the high-frequency limit can be rigorously defined in terms of the commutator of the current operators
\begin{equation} \label{eq:Hall_high_freq}
\sigma_H(\omega) \approx \frac{i}{N\omega^2} \expval{[ J_x, J_y]}.
\end{equation}
Note that this expression does not require the small-gap restriction. We can identify the sum in \eqref{eq:TROBterm} as the small-gap approximation of the expectation value in \eqref{eq:Hall_high_freq}. This expectation value also appears in a sum rule for the imaginary part of the Hall conductivity 
\begin{equation} \label{eq:sumrule}
\int_{-\infty}^\infty \omega \Im{\sigma_H(\omega)} \, \mathrm d \omega = -\frac{i\pi}{N} \expval{[ J_x, J_y]}.
\end{equation}
The expectation value in \eqref{eq:sumrule} is unlikely to be zero unless the Hall conductivity itself is vanishing at all frequencies. Further, for the  expectation value to be nonzero when both TROBs are vanishing implies that the leading contributions to \eqref{eq:Hall_high_freq} and~\eqref{eq:sumrule} must be fourth-order or higher in the pairing potential; for these higher-order contributions to be nonzero but the second-order contribution to vanish places extremely stringent conditions on both the pairing potential and the normal state Hamiltonian, which would not generically be satisfied. We thus consider it likely that vanishing TROBs imply a vanishing Hall conductivity in a clean system at arbitrary frequency and temperature.

\section{Example Calculations} \label{sec:examples}
To exemplify the role of the TROBs, here we present calculations for
two model systems.

\subsection{Strontium Ruthenate} \label{sec:sro}
Strontium ruthenate (Sr$_2$RuO$_4$) is a key candidate material for chiral superconductivity, with both muon spin relaxation \cite{luke_time-reversal_1998} and polar Kerr experiments \cite{kapitulnik_polar_2009,xia_high_2006} providing strong evidence of TRSB in its superconducting state. The theory of the anomalous Hall effect in Sr$_2$RuO$_4$ has been considered by several authors \cite{taylor_intrinsic_2012,gradhand_kerr_2013,wysokinski_intrinsic_2012}, all working on the assumption that it is a chiral $p$-wave superconductor \cite{ishida_knight_1998,duffy_polarized_2000,ishida_spin_2001}. Although recent experiments have thrown significant doubt on this proposition \cite{pustogow_constraints_2019,ishida_reduction_2020}, we nevertheless adopt this picture in order to make contact with the existing literature. We follow Ref.~\cite{taylor_intrinsic_2012} and adopt a minimal two-dimensional model of Sr$_2$RuO$_4$ involving the ruthenium $d_{xz}$ and $d_{yz}$ orbitals. Since the orbitals transform trivially under inversion and time-reversal, \eqref{alpha_beta_terms_same_partiy} enumerates the terms permitted in $H_0$. We take the following tight-binding model: $h_{00}=-t_1(\cos k_x + \cos k_y )-\mu$, $h_{10} = 2t_3\sin{k_x}\sin{k_y}$, $h_{30}=-t_2(\cos{k_x}-\cos{k_y})$, and isotropic spin orbit coupling $h_{23}=\lambda$; the $(2,1)$ and $(2,2)$ terms  do not appear in this two-dimensional model because $h_{21}$ and $h_{22}$ are odd under reflection about the $k_z=0$ mirror plane. Setting $t_1=1$, $\mu = 1$, $t_2=0.8$, $t_3=0.1$, and $\lambda=0.25$, this model qualitatively reproduces the observed $\alpha$ and $\beta$ Fermi surfaces~\cite{tamai_high-resolution_2019,rozbicki_spinorbit_2011,haverkort_strong_2008}.

We focus on pairing states belonging to $E_u$ irreducible representations of the $D_{4h}$ point group, which, being two dimensional, naturally occur in TRSB combinations. A general $E_u$ pairing state is given by $\tilde\Delta = \tilde\Delta_{03} \eta_0 \otimes \sigma_3 + \tilde\Delta_{13} \eta_1 \otimes \sigma_3 + \tilde\Delta_{20} \eta_2 \otimes \sigma_0  + \tilde\Delta_{33} \eta_3 \otimes \sigma_3$, where each $\tilde\Delta_{\alpha\beta}$ is an odd chiral function of momentum. The TROBs evaluate as 
\begin{widetext}
	\begin{align}
		\mathrm{TROB}_1 & = 4 \left[
		  \Im\{\tilde\Delta_{13} \tilde\Delta_{33}^*\} \eta_2 \otimes \sigma_0
		+ \Im\{\tilde\Delta_{33} \tilde\Delta_{20}^*\} \eta_1 \otimes \sigma_3
		+ \Im\{\tilde\Delta_{20} \tilde\Delta_{13}^*\} \eta_3 \otimes \sigma_3
	\right], \label{eq:TROB1_sro}                                                                 \\
		\mathrm{TROB}_2 & = 4 \left[(
			  \hat h_{10}\Im\{\tilde\Delta_{20} \tilde\Delta_{33}^*\}
		 	+ \hat h_{30}\Im\{\tilde\Delta_{11} \tilde\Delta_{22}^*\}
		 	+ \hat h_{23}\Im\{\tilde\Delta_{33} \tilde\Delta_{13}^*\} ) \eta_0 \otimes \sigma_3\right. \notag \\
		                & \phantom{=[} + (
			  \hat h_{30}\Im\{\tilde\Delta_{03} \tilde\Delta_{20}^*\} 
			+ \hat h_{23}\Im\{\tilde\Delta_{33} \tilde\Delta_{03}^*\}	) \eta_1 \otimes \sigma_3 
	+ (
			  \hat h_{10}\Im\{\tilde\Delta_{03} \tilde\Delta_{33}^*\} 
			+ \hat h_{30}\Im\{\tilde\Delta_{13} \tilde\Delta_{03}^*\}) \eta_2 \otimes \sigma_0 \notag \\
		                & \phantom{=[} \left.+ (
			  \hat h_{23}\Im\{\tilde\Delta_{03} \tilde\Delta_{13}^*\} 
			+ \hat h_{10}\Im\{\tilde\Delta_{20} \tilde\Delta_{03}^*\}) \eta_3 \otimes \sigma_3
	\right]. \label{eq:TROB2_sro}
	\end{align}
\end{widetext}
In general, both TROBs are non-zero, although they can each individually vanish depending on the details of a given model. Further, each term in \eqref{eq:TROB1_sro} and \eqref{eq:TROB2_sro} involves two pairing channels: although chiral pairing in a single channel is sufficient to break time-reversal symmetry, pairing in at least two channels is required for the appearance of an anomalous Hall conductivity.

\begin{figure*}
	\centering
	\hspace*{\fill}
	\includegraphics[width=0.32\linewidth]{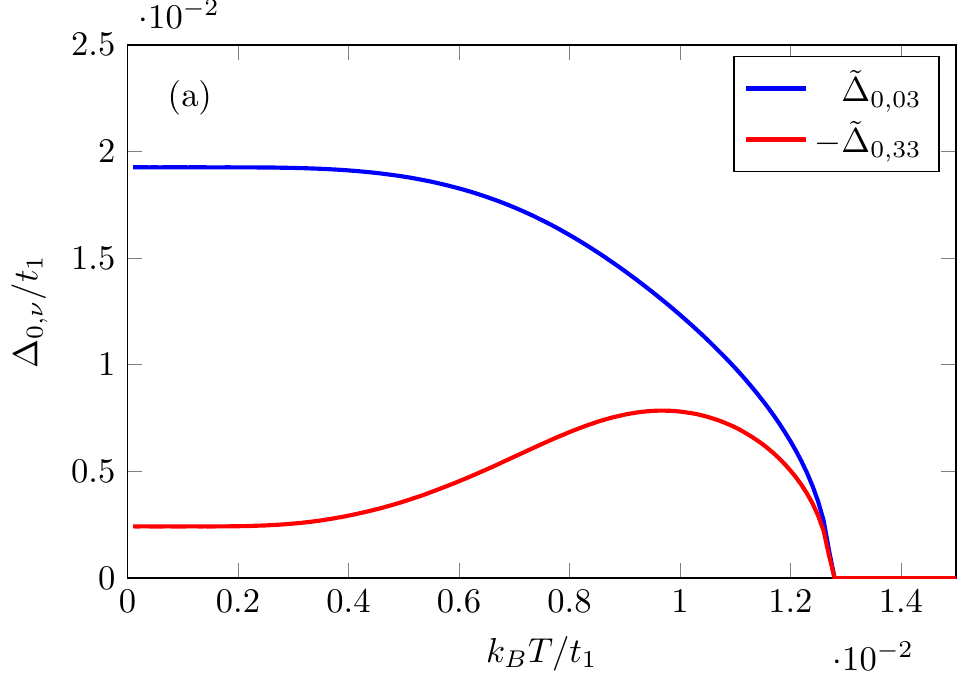}
	\hspace*{\fill}
	\includegraphics[width=0.32\linewidth]{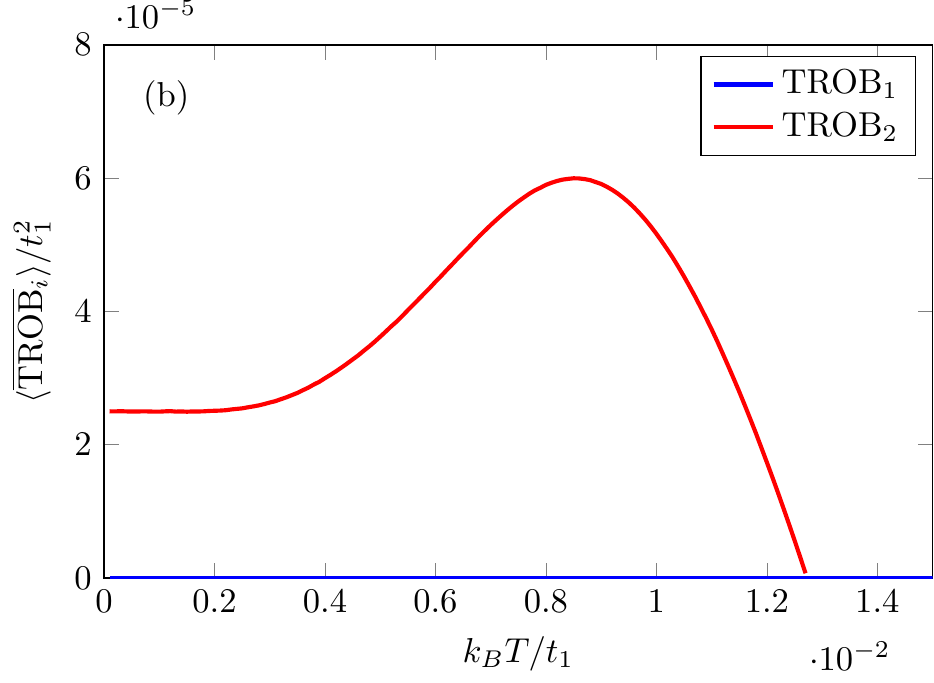}
	\hspace*{\fill}
	\includegraphics[width=0.32\linewidth]{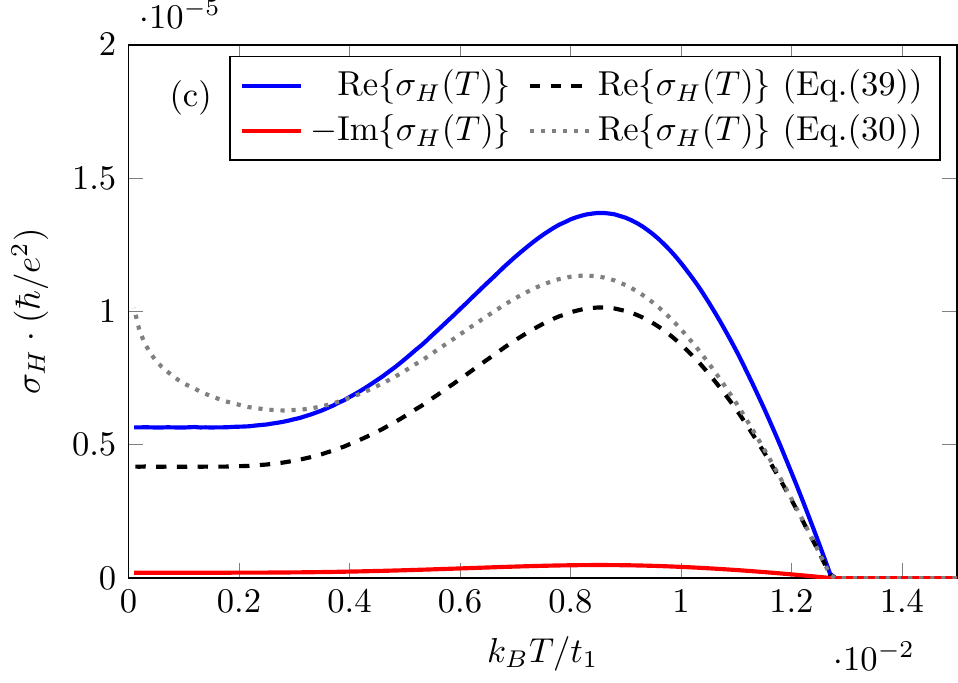}
	\hspace*{\fill}
	\caption{The pairing potential amplitudes (a), dimensionless TROB expectation values (b), and Hall conductivity at frequency $\omega=2t_1$ (c), calculated as a function of temperature for the model of strontium ruthenate discussed in the text. The Hall conductivity is shown both exactly (calculated using \eqref{eq:hall_def} and \eqref{eq:correlator_exact}) and in its approximate forms in the high-frequency and high-frequency, small-gap limits (using \eqref{eq:Hall_high_freq} and \eqref{eq:TROBterm} respectively). Calculated with $N=500\times500$ for (a) and (b), and $N=5000\times5000$ for (c). The positive infinitesimal in \eqref{eq:hall_def} was numerically approximated as $0^+=0.001$.}
	\label{fig:TROB_SRO}
\end{figure*}

As a specific model for further examination we take the intraorbital spin-triplet pairing state considered in Ref.~\cite{taylor_intrinsic_2012}
\begin{equation}
\tilde \Delta = \tilde\Delta_{03} \eta_0 \otimes \sigma_3 + \tilde\Delta_{33} \eta_3 \otimes \sigma_3,
\end{equation}
with the $p$-wave form  factors $\tilde\Delta_{03} = \tilde\Delta_{0,03} (\sin k_x + i \sin k_y)$ and $\tilde\Delta_{33} = \tilde\Delta_{0,33} (\sin k_x - i \sin k_y)$. Only $\mathrm{TROB}_2$ is nonzero for this system, evaluating to
\begin{equation}
\mathrm{TROB}_2 = 4\Im\{\tilde\Delta_{03}\tilde\Delta_{33}^*\} (\hat h_{10}  \eta_2 \otimes \sigma_0 - \hat h_{23} \eta_1 \otimes \sigma_3).
\end{equation}
Microscopically, $\mathrm{TROB}_2$ can be interpreted as an orbital- or spin-angular-momentum-polarized interorbital hopping term. 

To study the temperature-dependence of the TROB, we determine the pairing potential amplitudes $\tilde\Delta_{0,01}$ and $\tilde\Delta_{0,31}$ self-consistently. To this end, we introduce a phenomenological pairing interaction $V_{\nu,\nu'}$, which scatters a Cooper pair from channel $\nu'$ to channel $\nu$. The interaction strengths were taken to be $V_{03,03}=-0.2$, $V_{33,33}=-0.265$, and $V_{03,33}=V_{33,03}=0.03$. These values were chosen such that $\tilde\Delta_{0,33}$ exhibits a nonmonotonic temperature dependence, as shown in Fig.~\ref{fig:TROB_SRO}(a). The existence of TROB$_2$ in this state is confirmed by the thermal expectation value of the dimensionless quantity $\overline{\mathrm{TROB}}_i = \mathrm{TROB}_i/|\tilde\Delta_{0,03} \tilde\Delta_{0,33}|$, which is shown in Fig.~\ref{fig:TROB_SRO}(b). The nonmonotonic temperature dependence of the TROB is particularly helpful when comparing the approximate form of the Hall conductivity \eqref{eq:TROBterm} with the full calculation, which is shown in Fig.~\ref{fig:TROB_SRO}(c). We additionally include the prediction of the high-frequency limit \eqref{eq:Hall_high_freq}. The two approximate results are in excellent qualitative agreement with the full calculation, and also in reasonable quantitative agreement; the quantitative agreement improves at higher values of the frequency. Although this underscores the fact that the small-gap approximation captures the leading contribution to the Hall conductivity, we note that this approximation appears to break down as the temperature goes to zero.

\subsection{The honeycomb lattice}
An example of a TRSB even-parity superconductor with a nonzero anomalous Hall conductivity is provided by the nearest-neighbor chiral $d$-wave pairing state on the honeycomb lattice~\cite{BlackSchaffer_graphite_2007,Wu_2013}. The anomalous Hall conductivity of this system was analyzed in detail in~\cite{brydon_loop_2019}, which included the introduction of the TROB concept. In the notation used here, the honeycomb lattice model is written as 
\begin{equation}
H_0 = h_{00} \eta_0 \otimes \sigma_0 + h_{10} \eta_1 \otimes \sigma_0 + h_{20} \eta_2 \otimes \sigma_0,
\end{equation} 
where $\eta_\mu$ encodes the sublattice degree of freedom. The allowed terms are given by $h_{00}=\mu$, $h_{10}=-t(\cos{k_x}+2\cos\tfrac{1}{2}k_x\cos\tfrac{\sqrt{3}}{2}k_y)$, and $h_{20}=t(\sin{k_x}-2\sin\tfrac{1}{2}k_x\cos\tfrac{\sqrt{3}}{2}k_y)$, where $\mu$ is the chemical potential and $t$ is the nearest-neighbor hopping. The chiral $d$-wave pairing potential in the irreducible representation $E_{2g}$ of $D_{6h}$ is written 
\begin{equation}
\tilde \Delta = \tilde\Delta_{10} \eta_1 \otimes \sigma_0 + \tilde\Delta_{20} \eta_2 \otimes \sigma_0,
\end{equation}
with the amplitudes $\tilde\Delta_{10} = \Delta_{0}(\cos k_x - \cos\tfrac{k_x}{2}\cos\tfrac{\sqrt{3}}{2}k_y - i\sqrt{3}\sin\tfrac{k_x}{2}\sin\tfrac{\sqrt{3}k_y}{2})$ and $\tilde\Delta_{20} = -\Delta_{0}(\sin k_x + \sin\tfrac{k_x}{2}\cos\tfrac{\sqrt{3}k_y}{2} + i\sqrt{3}\cos\tfrac{k_x}{2}\sin\tfrac{\sqrt{3}k_y}{2})$. Only TROB$_{1}$ is nonzero for this model, taking the value
\begin{equation}
\text{TROB}_1 = 4\text{Im}\{ \tilde\Delta_{10}\tilde\Delta_{20}^\ast\}\eta_3\otimes\sigma_0.
\end{equation}
The form of TROB$_1$ is equivalent to the loop current term in Haldane's model of the anomalous quantum Hall effect, and the anomalous Hall conductivity of the model is directly related to this quantity~\cite{brydon_loop_2019}. The absence of TROB$_2$ is an artifact of the simplicity of the model: by including the symmetry-allowed Kane-Mele spin-orbit coupling~\cite{Kane_Mele_2005} $h_{33}\eta_3\otimes\sigma_3=\lambda \sin \tfrac{\sqrt{3}}{2}k_y(\cos\tfrac{3}{2}k_x-\cos\tfrac{\sqrt{3}}{2}k_y)\eta_3\otimes\sigma_3$ in the normal state Hamiltonian, we find
\begin{equation}
\mathrm{TROB}_2 = 4 \hat h_{33} \Im\{ \tilde\Delta_{10}^*\tilde\Delta_{20}\} \eta_0 \otimes \sigma_3,
\end{equation}
while TROB$_1$ is left unchanged. 

The TROB results for the honeycomb lattice are readily understood in terms of the band-pseudospin picture. In the absence of spin-orbit coupling, the pseudospin can be chosen as identical to the real spin, and so the interband pairing is purely singlet. Since the intraband singlet potentials have opposite sign, it follows that only TROB$_1$ is nonzero. The addition of the spin-orbit coupling introduces interband triplet pairing, and the second term in \eqref{eq:TROB2_even} for  TROB$_2$ is therefore nonzero.

\section{Conclusions and outlook}\label{sec:conclusions}
Motivated by the observation of the polar Kerr effect in chiral-superconductor candidate materials, in this manuscript we have considered the origin of the anomalous Hall conductivity in a general model of a two-band superconductor. Our analysis reveals a key role for two distinct gauge-invariant time-reversal-odd bilinear functions (TROBs) of the pairing potential: if at least one of these TROBs is present, we may generically expect that the anomalous Hall conductivity is nonzero. Although our result is rigorously valid in the high-frequency, small-gap limit, we argued that it applies more generally. The TROBs strongly constrain the form of the pairing potential which generates an anomalous Hall conductivity. Explicitly  calculating the TROBs in a pseudospin-band basis, we found that interband pairing generically gives rise to an anomalous Hall conductivity, but also demonstrated that purely-intraband nonunitary triplet pairing can make a contribution. Our general conclusions are illustrated with two specific examples of model chiral $p$-wave and $d$-wave superconductors. 

For the purposes of analytical clarity, our study has been restricted to a model with a twofold orbital degree of freedom. It nevertheless appears straightforward to generalize our analysis to systems with more orbital degrees of freedom. For example, more realistic models of the low-energy electronic structure of Sr$_2$RuO$_4$ typically involve at least three bands~\cite{wysokinski_intrinsic_2012,gradhand_kerr_2013,robbins_effect_2017,Zhang_SRO_2020,SuhMenke_SRO_2020}. We speculate that in a system with $n$ orbital degrees of freedom there will be $n$ distinct TROBs; this follows from the observation that the normal state Hamiltonian $H_0(\bk)$ of such a system obeys the characteristic polynomial
\begin{equation} \label{eq:charpoly}
\prod_{j=1}^n[H_0(\bk) - E_{\bk,j}] = 0
\end{equation}
where $E_{\bk,j}$ are the $n$ distinct doubly-degenerate eigenvalues of the normal state. This implies that we can define $n$ independent TROBs
\begin{equation}
\text{TROB}_j = \tilde{\Delta}\tilde{H}_0^{j-1}\tilde{\Delta}^\dagger - \tilde{\Delta}^\dagger\tilde{H}_0^{j-1}\tilde{\Delta},
\end{equation}
where $\tilde{H}_0$ is the traceless part of $H_0$. Due to the characteristic polynomial~\eqref{eq:charpoly}, TROBs defined in terms of powers of $H_0$ higher than $n-1$ can be expressed in terms of $\text{TROB}_{j=1,\ldots,n}$. We leave a detailed analysis of this situation to future work. 

Further generalization of our results should consider going beyond the small-gap limit, to explore the relevance of TROBs at all energy scales. Another direction for future work is to extend our argument to noncentrosymmetric systems, which could be relevant to time-reversal symmetry-breaking candidate superconductors LaNiC$_2$ \cite{hillier_evidence_2009} and twisted bilayer graphene~\cite{cao_unconventional_2018}.

\acknowledgments
This work was supported by the Marsden Fund Council from Government funding, managed by Royal Society Te Ap\={a}rangi.

\bibliography{article}

\begin{thebibliography}{47}%
\makeatletter
\providecommand \@ifxundefined [1]{%
 \@ifx{#1\undefined}
}%
\providecommand \@ifnum [1]{%
 \ifnum #1\expandafter \@firstoftwo
 \else \expandafter \@secondoftwo
 \fi
}%
\providecommand \@ifx [1]{%
 \ifx #1\expandafter \@firstoftwo
 \else \expandafter \@secondoftwo
 \fi
}%
\providecommand \natexlab [1]{#1}%
\providecommand \enquote  [1]{``#1''}%
\providecommand \bibnamefont  [1]{#1}%
\providecommand \bibfnamefont [1]{#1}%
\providecommand \citenamefont [1]{#1}%
\providecommand \href@noop [0]{\@secondoftwo}%
\providecommand \href [0]{\begingroup \@sanitize@url \@href}%
\providecommand \@href[1]{\@@startlink{#1}\@@href}%
\providecommand \@@href[1]{\endgroup#1\@@endlink}%
\providecommand \@sanitize@url [0]{\catcode `\\12\catcode `\$12\catcode
  `\&12\catcode `\#12\catcode `\^12\catcode `\_12\catcode `\%12\relax}%
\providecommand \@@startlink[1]{}%
\providecommand \@@endlink[0]{}%
\providecommand \url  [0]{\begingroup\@sanitize@url \@url }%
\providecommand \@url [1]{\endgroup\@href {#1}{\urlprefix }}%
\providecommand \urlprefix  [0]{URL }%
\providecommand \Eprint [0]{\href }%
\providecommand \doibase [0]{https://doi.org/}%
\providecommand \selectlanguage [0]{\@gobble}%
\providecommand \bibinfo  [0]{\@secondoftwo}%
\providecommand \bibfield  [0]{\@secondoftwo}%
\providecommand \translation [1]{[#1]}%
\providecommand \BibitemOpen [0]{}%
\providecommand \bibitemStop [0]{}%
\providecommand \bibitemNoStop [0]{.\EOS\space}%
\providecommand \EOS [0]{\spacefactor3000\relax}%
\providecommand \BibitemShut  [1]{\csname bibitem#1\endcsname}%
\let\auto@bib@innerbib\@empty
\bibitem [{\citenamefont {Kallin}\ and\ \citenamefont
  {Berlinsky}(2016)}]{kallin_chiral_2016}%
  \BibitemOpen
  \bibfield  {author} {\bibinfo {author} {\bibfnamefont {C.}~\bibnamefont
  {Kallin}}\ and\ \bibinfo {author} {\bibfnamefont {J.}~\bibnamefont
  {Berlinsky}},\ }\bibfield  {title} {\bibinfo {title} {Chiral
  superconductors},\ }\href {https://doi.org/10.1088/0034-4885/79/5/054502}
  {\bibfield  {journal} {\bibinfo  {journal} {Rep. Prog. Phys.}\ }\textbf
  {\bibinfo {volume} {79}},\ \bibinfo {pages} {054502} (\bibinfo {year}
  {2016})}\BibitemShut {NoStop}%
\bibitem [{\citenamefont {Mackenzie}\ and\ \citenamefont
  {Maeno}(2003)}]{mackenzie_superconductivity_2003}%
  \BibitemOpen
  \bibfield  {author} {\bibinfo {author} {\bibfnamefont {A.~P.}\ \bibnamefont
  {Mackenzie}}\ and\ \bibinfo {author} {\bibfnamefont {Y.}~\bibnamefont
  {Maeno}},\ }\bibfield  {title} {\bibinfo {title} {The superconductivity of
  $\mathrm{Sr}_2\mathrm{RuO}_4$ and the physics of spin-triplet pairing},\
  }\href {https://doi.org/10.1103/RevModPhys.75.657} {\bibfield  {journal}
  {\bibinfo  {journal} {Rev. Mod. Phys.}\ }\textbf {\bibinfo {volume} {75}},\
  \bibinfo {pages} {657} (\bibinfo {year} {2003})}\BibitemShut {NoStop}%
\bibitem [{\citenamefont {Norman}(1992)}]{norman_1992}%
  \BibitemOpen
  \bibfield  {author} {\bibinfo {author} {\bibfnamefont {M.}~\bibnamefont
  {Norman}},\ }\bibfield  {title} {\bibinfo {title} {What is the
  superconducting order parameter for {UPt$_3$}?},\ }\href
  {https://doi.org/10.1016/0921-4534(92)90692-6} {\bibfield  {journal}
  {\bibinfo  {journal} {Physica C: Superconductivity}\ }\textbf {\bibinfo
  {volume} {194}},\ \bibinfo {pages} {203 } (\bibinfo {year}
  {1992})}\BibitemShut {NoStop}%
\bibitem [{\citenamefont {Joynt}\ and\ \citenamefont
  {Taillefer}(2002)}]{joynt_2002}%
  \BibitemOpen
  \bibfield  {author} {\bibinfo {author} {\bibfnamefont {R.}~\bibnamefont
  {Joynt}}\ and\ \bibinfo {author} {\bibfnamefont {L.}~\bibnamefont
  {Taillefer}},\ }\bibfield  {title} {\bibinfo {title} {The superconducting
  phases of {UPt$_3$}},\ }\href {https://doi.org/10.1103/RevModPhys.74.235}
  {\bibfield  {journal} {\bibinfo  {journal} {Rev. Mod. Phys.}\ }\textbf
  {\bibinfo {volume} {74}},\ \bibinfo {pages} {235} (\bibinfo {year}
  {2002})}\BibitemShut {NoStop}%
\bibitem [{\citenamefont {Kasahara}\ \emph {et~al.}(2007)\citenamefont
  {Kasahara}, \citenamefont {Iwasawa}, \citenamefont {Shishido}, \citenamefont
  {Shibauchi}, \citenamefont {Behnia}, \citenamefont {Haga}, \citenamefont
  {Matsuda}, \citenamefont {Onuki}, \citenamefont {Sigrist},\ and\
  \citenamefont {Matsuda}}]{kasahara_exotic_2007}%
  \BibitemOpen
  \bibfield  {author} {\bibinfo {author} {\bibfnamefont {Y.}~\bibnamefont
  {Kasahara}}, \bibinfo {author} {\bibfnamefont {T.}~\bibnamefont {Iwasawa}},
  \bibinfo {author} {\bibfnamefont {H.}~\bibnamefont {Shishido}}, \bibinfo
  {author} {\bibfnamefont {T.}~\bibnamefont {Shibauchi}}, \bibinfo {author}
  {\bibfnamefont {K.}~\bibnamefont {Behnia}}, \bibinfo {author} {\bibfnamefont
  {Y.}~\bibnamefont {Haga}}, \bibinfo {author} {\bibfnamefont {T.~D.}\
  \bibnamefont {Matsuda}}, \bibinfo {author} {\bibfnamefont {Y.}~\bibnamefont
  {Onuki}}, \bibinfo {author} {\bibfnamefont {M.}~\bibnamefont {Sigrist}},\
  and\ \bibinfo {author} {\bibfnamefont {Y.}~\bibnamefont {Matsuda}},\
  }\bibfield  {title} {\bibinfo {title} {Exotic superconducting properties in
  the electron-hole-compensated heavy-fermion ``semimetal''
  {${\mathrm{URu}}_{2}{\mathrm{Si}}_{2}$}},\ }\href
  {https://doi.org/10.1103/PhysRevLett.99.116402} {\bibfield  {journal}
  {\bibinfo  {journal} {Phys. Rev. Lett.}\ }\textbf {\bibinfo {volume} {99}},\
  \bibinfo {pages} {116402} (\bibinfo {year} {2007})}\BibitemShut {NoStop}%
\bibitem [{\citenamefont {Cao}\ \emph {et~al.}(2018)\citenamefont {Cao},
  \citenamefont {Fatemi}, \citenamefont {Fang}, \citenamefont {Watanabe},
  \citenamefont {Taniguchi}, \citenamefont {Kaxiras},\ and\ \citenamefont
  {Jarillo-Herrero}}]{cao_unconventional_2018}%
  \BibitemOpen
  \bibfield  {author} {\bibinfo {author} {\bibfnamefont {Y.}~\bibnamefont
  {Cao}}, \bibinfo {author} {\bibfnamefont {V.}~\bibnamefont {Fatemi}},
  \bibinfo {author} {\bibfnamefont {S.}~\bibnamefont {Fang}}, \bibinfo {author}
  {\bibfnamefont {K.}~\bibnamefont {Watanabe}}, \bibinfo {author}
  {\bibfnamefont {T.}~\bibnamefont {Taniguchi}}, \bibinfo {author}
  {\bibfnamefont {E.}~\bibnamefont {Kaxiras}},\ and\ \bibinfo {author}
  {\bibfnamefont {P.}~\bibnamefont {Jarillo-Herrero}},\ }\bibfield  {title}
  {\bibinfo {title} {Unconventional superconductivity in magic-angle graphene
  superlattices},\ }\href {https://doi.org/10.1038/nature26160} {\bibfield
  {journal} {\bibinfo  {journal} {Nature}\ }\textbf {\bibinfo {volume} {556}},\
  \bibinfo {pages} {43} (\bibinfo {year} {2018})}\BibitemShut {NoStop}%
\bibitem [{\citenamefont {Argyres}(1955)}]{argyres_theory_1955}%
  \BibitemOpen
  \bibfield  {author} {\bibinfo {author} {\bibfnamefont {P.~N.}\ \bibnamefont
  {Argyres}},\ }\bibfield  {title} {\bibinfo {title} {Theory of the {Faraday}
  and {Kerr} {Effects} in {Ferromagnetics}},\ }\href
  {https://doi.org/10.1103/PhysRev.97.334} {\bibfield  {journal} {\bibinfo
  {journal} {Phys. Rev.}\ }\textbf {\bibinfo {volume} {97}},\ \bibinfo {pages}
  {334} (\bibinfo {year} {1955})}\BibitemShut {NoStop}%
\bibitem [{\citenamefont {Kapitulnik}\ \emph {et~al.}(2009)\citenamefont
  {Kapitulnik}, \citenamefont {Xia}, \citenamefont {Schemm},\ and\
  \citenamefont {Palevski}}]{kapitulnik_polar_2009}%
  \BibitemOpen
  \bibfield  {author} {\bibinfo {author} {\bibfnamefont {A.}~\bibnamefont
  {Kapitulnik}}, \bibinfo {author} {\bibfnamefont {J.}~\bibnamefont {Xia}},
  \bibinfo {author} {\bibfnamefont {E.}~\bibnamefont {Schemm}},\ and\ \bibinfo
  {author} {\bibfnamefont {A.}~\bibnamefont {Palevski}},\ }\bibfield  {title}
  {\bibinfo {title} {Polar {Kerr} effect as probe for time-reversal symmetry
  breaking in unconventional superconductors},\ }\href
  {https://doi.org/10.1088/1367-2630/11/5/055060} {\bibfield  {journal}
  {\bibinfo  {journal} {New J. Phys.}\ }\textbf {\bibinfo {volume} {11}},\
  \bibinfo {pages} {055060} (\bibinfo {year} {2009})}\BibitemShut {NoStop}%
\bibitem [{\citenamefont {Xia}\ \emph {et~al.}(2006)\citenamefont {Xia},
  \citenamefont {Maeno}, \citenamefont {Beyersdorf}, \citenamefont {Fejer},\
  and\ \citenamefont {Kapitulnik}}]{xia_high_2006}%
  \BibitemOpen
  \bibfield  {author} {\bibinfo {author} {\bibfnamefont {J.}~\bibnamefont
  {Xia}}, \bibinfo {author} {\bibfnamefont {Y.}~\bibnamefont {Maeno}}, \bibinfo
  {author} {\bibfnamefont {P.~T.}\ \bibnamefont {Beyersdorf}}, \bibinfo
  {author} {\bibfnamefont {M.~M.}\ \bibnamefont {Fejer}},\ and\ \bibinfo
  {author} {\bibfnamefont {A.}~\bibnamefont {Kapitulnik}},\ }\bibfield  {title}
  {\bibinfo {title} {High resolution polar {Kerr} effect measurements of
  $\mathrm{Sr}_2\mathrm{RuO}_4$: Evidence for broken time-reversal symmetry in
  the superconducting state},\ }\href
  {https://doi.org/10.1103/PhysRevLett.97.167002} {\bibfield  {journal}
  {\bibinfo  {journal} {Phys. Rev. Lett.}\ }\textbf {\bibinfo {volume} {97}},\
  \bibinfo {pages} {167002} (\bibinfo {year} {2006})}\BibitemShut {NoStop}%
\bibitem [{\citenamefont {Schemm}\ \emph {et~al.}(2014)\citenamefont {Schemm},
  \citenamefont {Gannon}, \citenamefont {Wishne}, \citenamefont {Halperin},\
  and\ \citenamefont {Kapitulnik}}]{schemm_observation_2014}%
  \BibitemOpen
  \bibfield  {author} {\bibinfo {author} {\bibfnamefont {E.~R.}\ \bibnamefont
  {Schemm}}, \bibinfo {author} {\bibfnamefont {W.~J.}\ \bibnamefont {Gannon}},
  \bibinfo {author} {\bibfnamefont {C.~M.}\ \bibnamefont {Wishne}}, \bibinfo
  {author} {\bibfnamefont {W.~P.}\ \bibnamefont {Halperin}},\ and\ \bibinfo
  {author} {\bibfnamefont {A.}~\bibnamefont {Kapitulnik}},\ }\bibfield  {title}
  {\bibinfo {title} {Observation of broken time-reversal symmetry in the
  heavy-fermion superconductor {UPt$_3$}},\ }\href
  {https://doi.org/10.1126/science.1248552} {\bibfield  {journal} {\bibinfo
  {journal} {Science}\ }\textbf {\bibinfo {volume} {345}},\ \bibinfo {pages}
  {190} (\bibinfo {year} {2014})}\BibitemShut {NoStop}%
\bibitem [{\citenamefont {Schemm}\ \emph {et~al.}(2015)\citenamefont {Schemm},
  \citenamefont {Baumbach}, \citenamefont {Tobash}, \citenamefont {Ronning},
  \citenamefont {Bauer},\ and\ \citenamefont
  {Kapitulnik}}]{schemm_evidence_2015}%
  \BibitemOpen
  \bibfield  {author} {\bibinfo {author} {\bibfnamefont {E.~R.}\ \bibnamefont
  {Schemm}}, \bibinfo {author} {\bibfnamefont {R.~E.}\ \bibnamefont
  {Baumbach}}, \bibinfo {author} {\bibfnamefont {P.~H.}\ \bibnamefont
  {Tobash}}, \bibinfo {author} {\bibfnamefont {F.}~\bibnamefont {Ronning}},
  \bibinfo {author} {\bibfnamefont {E.~D.}\ \bibnamefont {Bauer}},\ and\
  \bibinfo {author} {\bibfnamefont {A.}~\bibnamefont {Kapitulnik}},\ }\bibfield
   {title} {\bibinfo {title} {Evidence for broken time-reversal symmetry in the
  superconducting phase of {${\mathrm{URu}}_{2}{\mathrm{Si}}_{2}$}},\ }\href
  {https://doi.org/10.1103/PhysRevB.91.140506} {\bibfield  {journal} {\bibinfo
  {journal} {Phys. Rev. B}\ }\textbf {\bibinfo {volume} {91}},\ \bibinfo
  {pages} {140506} (\bibinfo {year} {2015})}\BibitemShut {NoStop}%
\bibitem [{\citenamefont {Gong}\ \emph {et~al.}(2017)\citenamefont {Gong},
  \citenamefont {Kargarian}, \citenamefont {Stern}, \citenamefont {Yue},
  \citenamefont {Zhou}, \citenamefont {Jin}, \citenamefont {Galitski},
  \citenamefont {Yakovenko},\ and\ \citenamefont
  {Xia}}]{gong_time-reversal_2017}%
  \BibitemOpen
  \bibfield  {author} {\bibinfo {author} {\bibfnamefont {X.}~\bibnamefont
  {Gong}}, \bibinfo {author} {\bibfnamefont {M.}~\bibnamefont {Kargarian}},
  \bibinfo {author} {\bibfnamefont {A.}~\bibnamefont {Stern}}, \bibinfo
  {author} {\bibfnamefont {D.}~\bibnamefont {Yue}}, \bibinfo {author}
  {\bibfnamefont {H.}~\bibnamefont {Zhou}}, \bibinfo {author} {\bibfnamefont
  {X.}~\bibnamefont {Jin}}, \bibinfo {author} {\bibfnamefont {V.~M.}\
  \bibnamefont {Galitski}}, \bibinfo {author} {\bibfnamefont {V.~M.}\
  \bibnamefont {Yakovenko}},\ and\ \bibinfo {author} {\bibfnamefont
  {J.}~\bibnamefont {Xia}},\ }\bibfield  {title} {\bibinfo {title}
  {Time-reversal symmetry-breaking superconductivity in epitaxial
  bismuth/nickel bilayers},\ }\href {https://doi.org/10.1126/sciadv.1602579}
  {\bibfield  {journal} {\bibinfo  {journal} {Sci. Adv.}\ }\textbf {\bibinfo
  {volume} {3}},\ \bibinfo {pages} {1602579} (\bibinfo {year}
  {2017})}\BibitemShut {NoStop}%
\bibitem [{\citenamefont {Levenson-Falk}\ \emph {et~al.}(2018)\citenamefont
  {Levenson-Falk}, \citenamefont {Schemm}, \citenamefont {Aoki}, \citenamefont
  {Maple},\ and\ \citenamefont {Kapitulnik}}]{LevensonFalk_2018}%
  \BibitemOpen
  \bibfield  {author} {\bibinfo {author} {\bibfnamefont {E.~M.}\ \bibnamefont
  {Levenson-Falk}}, \bibinfo {author} {\bibfnamefont {E.~R.}\ \bibnamefont
  {Schemm}}, \bibinfo {author} {\bibfnamefont {Y.}~\bibnamefont {Aoki}},
  \bibinfo {author} {\bibfnamefont {M.~B.}\ \bibnamefont {Maple}},\ and\
  \bibinfo {author} {\bibfnamefont {A.}~\bibnamefont {Kapitulnik}},\ }\bibfield
   {title} {\bibinfo {title} {Polar kerr effect from time-reversal symmetry
  breaking in the heavy-fermion superconductor
  $\mathrm{PrOs}_{4}\mathrm{Sb}_{12}$},\ }\href
  {https://doi.org/10.1103/PhysRevLett.120.187004} {\bibfield  {journal}
  {\bibinfo  {journal} {Phys. Rev. Lett.}\ }\textbf {\bibinfo {volume} {120}},\
  \bibinfo {pages} {187004} (\bibinfo {year} {2018})}\BibitemShut {NoStop}%
\bibitem [{\citenamefont {Hayes}\ \emph {et~al.}(2020)\citenamefont {Hayes},
  \citenamefont {Wei}, \citenamefont {Metz}, \citenamefont {Zhang},
  \citenamefont {Eo}, \citenamefont {Ran}, \citenamefont {Saha}, \citenamefont
  {Collini}, \citenamefont {Butch}, \citenamefont {Agterberg}, \citenamefont
  {Kapitulnik},\ and\ \citenamefont {Paglione}}]{Hayes_UTe2}%
  \BibitemOpen
  \bibfield  {author} {\bibinfo {author} {\bibfnamefont {I.~M.}\ \bibnamefont
  {Hayes}}, \bibinfo {author} {\bibfnamefont {D.~S.}\ \bibnamefont {Wei}},
  \bibinfo {author} {\bibfnamefont {T.}~\bibnamefont {Metz}}, \bibinfo {author}
  {\bibfnamefont {J.}~\bibnamefont {Zhang}}, \bibinfo {author} {\bibfnamefont
  {Y.~S.}\ \bibnamefont {Eo}}, \bibinfo {author} {\bibfnamefont
  {S.}~\bibnamefont {Ran}}, \bibinfo {author} {\bibfnamefont {S.~R.}\
  \bibnamefont {Saha}}, \bibinfo {author} {\bibfnamefont {J.}~\bibnamefont
  {Collini}}, \bibinfo {author} {\bibfnamefont {N.~P.}\ \bibnamefont {Butch}},
  \bibinfo {author} {\bibfnamefont {D.~F.}\ \bibnamefont {Agterberg}}, \bibinfo
  {author} {\bibfnamefont {A.}~\bibnamefont {Kapitulnik}},\ and\ \bibinfo
  {author} {\bibfnamefont {J.}~\bibnamefont {Paglione}},\ }\href@noop {}
  {\bibinfo {title} {Weyl superconductivity in $\mathrm{UTe}_2$}} (\bibinfo
  {year} {2020}),\ \Eprint {https://arxiv.org/abs/2002.02539} {arXiv:2002.02539
  [cond-mat.str-el]} \BibitemShut {NoStop}%
\bibitem [{\citenamefont {Taylor}\ and\ \citenamefont
  {Kallin}(2012)}]{taylor_intrinsic_2012}%
  \BibitemOpen
  \bibfield  {author} {\bibinfo {author} {\bibfnamefont {E.}~\bibnamefont
  {Taylor}}\ and\ \bibinfo {author} {\bibfnamefont {C.}~\bibnamefont
  {Kallin}},\ }\bibfield  {title} {\bibinfo {title} {Intrinsic {Hall} effect in
  a multiband chiral superconductor in the absence of an external magnetic
  field},\ }\href {https://doi.org/10.1103/PhysRevLett.108.157001} {\bibfield
  {journal} {\bibinfo  {journal} {Phys. Rev. Lett.}\ }\textbf {\bibinfo
  {volume} {108}},\ \bibinfo {pages} {157001} (\bibinfo {year}
  {2012})}\BibitemShut {NoStop}%
\bibitem [{\citenamefont {K{\"o}nig}\ and\ \citenamefont
  {Levchenko}(2017)}]{konig_kerr_2017}%
  \BibitemOpen
  \bibfield  {author} {\bibinfo {author} {\bibfnamefont {E.~J.}\ \bibnamefont
  {K{\"o}nig}}\ and\ \bibinfo {author} {\bibfnamefont {A.}~\bibnamefont
  {Levchenko}},\ }\bibfield  {title} {\bibinfo {title} {Kerr effect from
  diffractive skew scattering in chiral $p_x\pm ip_y$ superconductors},\ }\href
  {https://doi.org/10.1103/PhysRevLett.118.027001} {\bibfield  {journal}
  {\bibinfo  {journal} {Phys. Rev. Lett.}\ }\textbf {\bibinfo {volume} {118}},\
  \bibinfo {pages} {027001} (\bibinfo {year} {2017})}\BibitemShut {NoStop}%
\bibitem [{\citenamefont {Goryo}(2008)}]{goryo_impurity-induced_2008}%
  \BibitemOpen
  \bibfield  {author} {\bibinfo {author} {\bibfnamefont {J.}~\bibnamefont
  {Goryo}},\ }\bibfield  {title} {\bibinfo {title} {Impurity-induced polar
  {Kerr} effect in a chiral $p$-wave superconductor},\ }\href
  {https://doi.org/10.1103/PhysRevB.78.060501} {\bibfield  {journal} {\bibinfo
  {journal} {Phys. Rev. B}\ }\textbf {\bibinfo {volume} {78}},\ \bibinfo
  {pages} {060501} (\bibinfo {year} {2008})}\BibitemShut {NoStop}%
\bibitem [{\citenamefont {Lutchyn}\ \emph {et~al.}(2009)\citenamefont
  {Lutchyn}, \citenamefont {Nagornykh},\ and\ \citenamefont
  {Yakovenko}}]{lutchyn_frequency_2009}%
  \BibitemOpen
  \bibfield  {author} {\bibinfo {author} {\bibfnamefont {R.~M.}\ \bibnamefont
  {Lutchyn}}, \bibinfo {author} {\bibfnamefont {P.}~\bibnamefont {Nagornykh}},\
  and\ \bibinfo {author} {\bibfnamefont {V.~M.}\ \bibnamefont {Yakovenko}},\
  }\bibfield  {title} {\bibinfo {title} {Frequency and temperature dependence
  of the anomalous ac {Hall} conductivity in a chiral $p_x+ip_y$ superconductor
  with impurities},\ }\href {https://doi.org/10.1103/PhysRevB.80.104508}
  {\bibfield  {journal} {\bibinfo  {journal} {Phys. Rev. B}\ }\textbf {\bibinfo
  {volume} {80}},\ \bibinfo {pages} {104508} (\bibinfo {year}
  {2009})}\BibitemShut {NoStop}%
\bibitem [{\citenamefont {Li}\ \emph {et~al.}(2015)\citenamefont {Li},
  \citenamefont {Andreev},\ and\ \citenamefont {Spivak}}]{li_anomalous_2015}%
  \BibitemOpen
  \bibfield  {author} {\bibinfo {author} {\bibfnamefont {S.}~\bibnamefont
  {Li}}, \bibinfo {author} {\bibfnamefont {A.~V.}\ \bibnamefont {Andreev}},\
  and\ \bibinfo {author} {\bibfnamefont {B.~Z.}\ \bibnamefont {Spivak}},\
  }\bibfield  {title} {\bibinfo {title} {Anomalous transport phenomena in
  ${p}_{x}+i{p}_{y}$ superconductors},\ }\href
  {https://doi.org/10.1103/PhysRevB.92.100506} {\bibfield  {journal} {\bibinfo
  {journal} {Phys. Rev. B}\ }\textbf {\bibinfo {volume} {92}},\ \bibinfo
  {pages} {100506} (\bibinfo {year} {2015})}\BibitemShut {NoStop}%
\bibitem [{\citenamefont {Li}\ \emph {et~al.}(2019)\citenamefont {Li},
  \citenamefont {Wang},\ and\ \citenamefont {Huang}}]{Zhang_AHE_2019}%
  \BibitemOpen
  \bibfield  {author} {\bibinfo {author} {\bibfnamefont {Y.}~\bibnamefont
  {Li}}, \bibinfo {author} {\bibfnamefont {Z.}~\bibnamefont {Wang}},\ and\
  \bibinfo {author} {\bibfnamefont {W.}~\bibnamefont {Huang}},\ }\href@noop {}
  {\bibinfo {title} {Anomalous {Hall} effect in chiral superconductors from
  impurity superlattices}} (\bibinfo {year} {2019}),\ \Eprint
  {https://arxiv.org/abs/1909.08012} {arXiv:1909.08012 [cond-mat.supr-con]}
  \BibitemShut {NoStop}%
\bibitem [{\citenamefont {Wysoki{\'n}ski}\ \emph {et~al.}(2012)\citenamefont
  {Wysoki{\'n}ski}, \citenamefont {Annett},\ and\ \citenamefont
  {Gy{\"o}rffy}}]{wysokinski_intrinsic_2012}%
  \BibitemOpen
  \bibfield  {author} {\bibinfo {author} {\bibfnamefont {K.~I.}\ \bibnamefont
  {Wysoki{\'n}ski}}, \bibinfo {author} {\bibfnamefont {J.~F.}\ \bibnamefont
  {Annett}},\ and\ \bibinfo {author} {\bibfnamefont {B.~L.}\ \bibnamefont
  {Gy{\"o}rffy}},\ }\bibfield  {title} {\bibinfo {title} {Intrinsic optical
  dichroism in the chiral superconducting state of
  $\mathrm{Sr}_2\mathrm{RuO}_4$},\ }\href
  {https://doi.org/10.1103/PhysRevLett.108.077004} {\bibfield  {journal}
  {\bibinfo  {journal} {Phys. Rev. Lett.}\ }\textbf {\bibinfo {volume} {108}},\
  \bibinfo {pages} {077004} (\bibinfo {year} {2012})}\BibitemShut {NoStop}%
\bibitem [{\citenamefont {Gradhand}\ \emph {et~al.}(2013)\citenamefont
  {Gradhand}, \citenamefont {Wysokinski}, \citenamefont {Annett},\ and\
  \citenamefont {Gy\"orffy}}]{gradhand_kerr_2013}%
  \BibitemOpen
  \bibfield  {author} {\bibinfo {author} {\bibfnamefont {M.}~\bibnamefont
  {Gradhand}}, \bibinfo {author} {\bibfnamefont {K.~I.}\ \bibnamefont
  {Wysokinski}}, \bibinfo {author} {\bibfnamefont {J.~F.}\ \bibnamefont
  {Annett}},\ and\ \bibinfo {author} {\bibfnamefont {B.~L.}\ \bibnamefont
  {Gy\"orffy}},\ }\bibfield  {title} {\bibinfo {title} {Kerr rotation in the
  unconventional superconductor {Sr${}_{2}$RuO${}_{4}$}},\ }\href
  {https://doi.org/10.1103/PhysRevB.88.094504} {\bibfield  {journal} {\bibinfo
  {journal} {Phys. Rev. B}\ }\textbf {\bibinfo {volume} {88}},\ \bibinfo
  {pages} {094504} (\bibinfo {year} {2013})}\BibitemShut {NoStop}%
\bibitem [{\citenamefont {Wang}\ \emph {et~al.}(2017)\citenamefont {Wang},
  \citenamefont {Berlinsky}, \citenamefont {Zwicknagl},\ and\ \citenamefont
  {Kallin}}]{wang_intrinsic_2017}%
  \BibitemOpen
  \bibfield  {author} {\bibinfo {author} {\bibfnamefont {Z.}~\bibnamefont
  {Wang}}, \bibinfo {author} {\bibfnamefont {J.}~\bibnamefont {Berlinsky}},
  \bibinfo {author} {\bibfnamefont {G.}~\bibnamefont {Zwicknagl}},\ and\
  \bibinfo {author} {\bibfnamefont {C.}~\bibnamefont {Kallin}},\ }\bibfield
  {title} {\bibinfo {title} {Intrinsic ac anomalous {Hall} effect of
  nonsymmorphic chiral superconductors with an application to
  $\mathrm{UPt}_3$},\ }\href {https://doi.org/10.1103/PhysRevB.96.174511}
  {\bibfield  {journal} {\bibinfo  {journal} {Phys. Rev. B}\ }\textbf {\bibinfo
  {volume} {96}},\ \bibinfo {pages} {174511} (\bibinfo {year}
  {2017})}\BibitemShut {NoStop}%
\bibitem [{\citenamefont {Robbins}\ \emph {et~al.}(2017)\citenamefont
  {Robbins}, \citenamefont {Annett},\ and\ \citenamefont
  {Gradhand}}]{robbins_effect_2017}%
  \BibitemOpen
  \bibfield  {author} {\bibinfo {author} {\bibfnamefont {J.}~\bibnamefont
  {Robbins}}, \bibinfo {author} {\bibfnamefont {J.~F.}\ \bibnamefont
  {Annett}},\ and\ \bibinfo {author} {\bibfnamefont {M.}~\bibnamefont
  {Gradhand}},\ }\bibfield  {title} {\bibinfo {title} {Effect of spin-orbit
  coupling on the polar {Kerr} effect in
  {${\mathrm{Sr}}_{2}{\mathrm{RuO}}_{4}$}},\ }\href
  {https://doi.org/10.1103/PhysRevB.96.144503} {\bibfield  {journal} {\bibinfo
  {journal} {Phys. Rev. B}\ }\textbf {\bibinfo {volume} {96}},\ \bibinfo
  {pages} {144503} (\bibinfo {year} {2017})}\BibitemShut {NoStop}%
\bibitem [{\citenamefont {Robert}\ and\ \citenamefont
  {Wen-Chin}(2017)}]{joynt_superconductivity_2017}%
  \BibitemOpen
  \bibfield  {author} {\bibinfo {author} {\bibfnamefont {J.}~\bibnamefont
  {Robert}}\ and\ \bibinfo {author} {\bibfnamefont {W.}~\bibnamefont
  {Wen-Chin}},\ }\bibfield  {title} {\bibinfo {title} {Superconductivity in
  empty bands and multiple order parameter chirality},\ }\href
  {https://doi.org/10.1038/s41598-017-13426-9} {\bibfield  {journal} {\bibinfo
  {journal} {Sci. Rep.}\ }\textbf {\bibinfo {volume} {7}},\ \bibinfo {pages}
  {12968} (\bibinfo {year} {2017})}\BibitemShut {NoStop}%
\bibitem [{\citenamefont {Triola}\ and\ \citenamefont
  {Black-Schaffer}(2018)}]{Triola_2018}%
  \BibitemOpen
  \bibfield  {author} {\bibinfo {author} {\bibfnamefont {C.}~\bibnamefont
  {Triola}}\ and\ \bibinfo {author} {\bibfnamefont {A.~M.}\ \bibnamefont
  {Black-Schaffer}},\ }\bibfield  {title} {\bibinfo {title} {Odd-frequency
  pairing and {Kerr} effect in the heavy-fermion superconductor
  $\mathrm{UPt}_{3}$},\ }\href {https://doi.org/10.1103/PhysRevB.97.064505}
  {\bibfield  {journal} {\bibinfo  {journal} {Phys. Rev. B}\ }\textbf {\bibinfo
  {volume} {97}},\ \bibinfo {pages} {064505} (\bibinfo {year}
  {2018})}\BibitemShut {NoStop}%
\bibitem [{\citenamefont {Brydon}\ \emph {et~al.}(2019)\citenamefont {Brydon},
  \citenamefont {Abergel}, \citenamefont {Agterberg},\ and\ \citenamefont
  {Yakovenko}}]{brydon_loop_2019}%
  \BibitemOpen
  \bibfield  {author} {\bibinfo {author} {\bibfnamefont {P.~M.~R.}\
  \bibnamefont {Brydon}}, \bibinfo {author} {\bibfnamefont {D.~S.~L.}\
  \bibnamefont {Abergel}}, \bibinfo {author} {\bibfnamefont {D.~F.}\
  \bibnamefont {Agterberg}},\ and\ \bibinfo {author} {\bibfnamefont {V.~M.}\
  \bibnamefont {Yakovenko}},\ }\bibfield  {title} {\bibinfo {title} {Loop
  currents and anomalous {Hall} effect from time-reversal symmetry-breaking
  superconductivity on the honeycomb lattice},\ }\href
  {https://doi.org/10.1103/PhysRevX.9.031025} {\bibfield  {journal} {\bibinfo
  {journal} {Phys. Rev. X}\ }\textbf {\bibinfo {volume} {9}},\ \bibinfo {pages}
  {031025} (\bibinfo {year} {2019})}\BibitemShut {NoStop}%
\bibitem [{\citenamefont {Zhang}\ \emph {et~al.}(2020)\citenamefont {Zhang},
  \citenamefont {Li}, \citenamefont {Huang},\ and\ \citenamefont
  {Zhang}}]{Zhang_SRO_2020}%
  \BibitemOpen
  \bibfield  {author} {\bibinfo {author} {\bibfnamefont {J.-L.}\ \bibnamefont
  {Zhang}}, \bibinfo {author} {\bibfnamefont {Y.}~\bibnamefont {Li}}, \bibinfo
  {author} {\bibfnamefont {W.}~\bibnamefont {Huang}},\ and\ \bibinfo {author}
  {\bibfnamefont {F.-C.}\ \bibnamefont {Zhang}},\ }\href@noop {} {\bibinfo
  {title} {Hidden anomalous {Hall} effect in $\mathrm{Sr}_{2}\mathrm{RuO}_{4}$
  with chiral superconductivity dominated by the $\mathrm{Ru}$ $d_{xy}$
  orbital}} (\bibinfo {year} {2020}),\ \Eprint
  {https://arxiv.org/abs/2009.00034} {arXiv:2009.00034 [cond-mat.supr-con]}
  \BibitemShut {NoStop}%
\bibitem [{\citenamefont {Yip}(2016)}]{yip_pseudospin_2016}%
  \BibitemOpen
  \bibfield  {author} {\bibinfo {author} {\bibfnamefont {S.}~\bibnamefont
  {Yip}},\ }\href@noop {} {\bibinfo {title} {Pseudospin bases for a model of
  {Cu:Bi$_2$Se$_3$}}} (\bibinfo {year} {2016}),\ \Eprint
  {https://arxiv.org/abs/1609.04152} {arXiv:1609.04152 [cond-mat.supr-con]}
  \BibitemShut {NoStop}%
\bibitem [{\citenamefont {Venderbos}\ \emph {et~al.}(2016)\citenamefont
  {Venderbos}, \citenamefont {Kozii},\ and\ \citenamefont
  {Fu}}]{venderbos_odd-parity_2016}%
  \BibitemOpen
  \bibfield  {author} {\bibinfo {author} {\bibfnamefont {J.~W.~F.}\
  \bibnamefont {Venderbos}}, \bibinfo {author} {\bibfnamefont {V.}~\bibnamefont
  {Kozii}},\ and\ \bibinfo {author} {\bibfnamefont {L.}~\bibnamefont {Fu}},\
  }\bibfield  {title} {\bibinfo {title} {Odd-parity superconductors with
  two-component order parameters: Nematic and chiral, full gap, and {Majorana}
  node},\ }\href {https://doi.org/10.1103/PhysRevB.94.180504} {\bibfield
  {journal} {\bibinfo  {journal} {Phys. Rev. B}\ }\textbf {\bibinfo {volume}
  {94}},\ \bibinfo {pages} {180504} (\bibinfo {year} {2016})}\BibitemShut
  {NoStop}%
\bibitem [{\citenamefont {Mahan}(1980)}]{mahan_many-particle_1980}%
  \BibitemOpen
  \bibfield  {author} {\bibinfo {author} {\bibfnamefont {G.~D.}\ \bibnamefont
  {Mahan}},\ }\href@noop {} {\emph {\bibinfo {title} {Many-{Particle}
  {Physics}}}},\ \bibinfo {edition} {3rd}\ ed.\ (\bibinfo {year}
  {1980})\BibitemShut {NoStop}%
\bibitem [{\citenamefont {Ramires}\ \emph {et~al.}(2018)\citenamefont
  {Ramires}, \citenamefont {Agterberg},\ and\ \citenamefont
  {Sigrist}}]{ramires_tailoring_2018}%
  \BibitemOpen
  \bibfield  {author} {\bibinfo {author} {\bibfnamefont {A.}~\bibnamefont
  {Ramires}}, \bibinfo {author} {\bibfnamefont {D.~F.}\ \bibnamefont
  {Agterberg}},\ and\ \bibinfo {author} {\bibfnamefont {M.}~\bibnamefont
  {Sigrist}},\ }\bibfield  {title} {\bibinfo {title} {Tailoring {$T_c$} by
  symmetry principles: {The} concept of superconducting fitness},\ }\href
  {https://doi.org/10.1103/PhysRevB.98.024501} {\bibfield  {journal} {\bibinfo
  {journal} {Phys. Rev. B}\ }\textbf {\bibinfo {volume} {98}},\ \bibinfo
  {pages} {024501} (\bibinfo {year} {2018})}\BibitemShut {NoStop}%
\bibitem [{\citenamefont {Brydon}\ \emph {et~al.}(2018)\citenamefont {Brydon},
  \citenamefont {Agterberg}, \citenamefont {Menke},\ and\ \citenamefont
  {Timm}}]{brydon_bfs_2018}%
  \BibitemOpen
  \bibfield  {author} {\bibinfo {author} {\bibfnamefont {P.~M.~R.}\
  \bibnamefont {Brydon}}, \bibinfo {author} {\bibfnamefont {D.~F.}\
  \bibnamefont {Agterberg}}, \bibinfo {author} {\bibfnamefont {H.}~\bibnamefont
  {Menke}},\ and\ \bibinfo {author} {\bibfnamefont {C.}~\bibnamefont {Timm}},\
  }\bibfield  {title} {\bibinfo {title} {Bogoliubov fermi surfaces: General
  theory, magnetic order, and topology},\ }\href
  {https://doi.org/10.1103/PhysRevB.98.224509} {\bibfield  {journal} {\bibinfo
  {journal} {Phys. Rev. B}\ }\textbf {\bibinfo {volume} {98}},\ \bibinfo
  {pages} {224509} (\bibinfo {year} {2018})}\BibitemShut {NoStop}%
\bibitem [{\citenamefont {Luke}\ \emph {et~al.}(1998)\citenamefont {Luke},
  \citenamefont {Fudamoto}, \citenamefont {Kojima}, \citenamefont {Larkin},
  \citenamefont {Merrin}, \citenamefont {Nachumi}, \citenamefont {Uemura},
  \citenamefont {Maeno}, \citenamefont {Mao}, \citenamefont {Mori},
  \citenamefont {Nakamura},\ and\ \citenamefont
  {Sigrist}}]{luke_time-reversal_1998}%
  \BibitemOpen
  \bibfield  {author} {\bibinfo {author} {\bibfnamefont {G.~M.}\ \bibnamefont
  {Luke}}, \bibinfo {author} {\bibfnamefont {Y.}~\bibnamefont {Fudamoto}},
  \bibinfo {author} {\bibfnamefont {K.~M.}\ \bibnamefont {Kojima}}, \bibinfo
  {author} {\bibfnamefont {M.~I.}\ \bibnamefont {Larkin}}, \bibinfo {author}
  {\bibfnamefont {J.}~\bibnamefont {Merrin}}, \bibinfo {author} {\bibfnamefont
  {B.}~\bibnamefont {Nachumi}}, \bibinfo {author} {\bibfnamefont {Y.~J.}\
  \bibnamefont {Uemura}}, \bibinfo {author} {\bibfnamefont {Y.}~\bibnamefont
  {Maeno}}, \bibinfo {author} {\bibfnamefont {Z.~Q.}\ \bibnamefont {Mao}},
  \bibinfo {author} {\bibfnamefont {Y.}~\bibnamefont {Mori}}, \bibinfo {author}
  {\bibfnamefont {H.}~\bibnamefont {Nakamura}},\ and\ \bibinfo {author}
  {\bibfnamefont {M.}~\bibnamefont {Sigrist}},\ }\bibfield  {title} {\bibinfo
  {title} {Time-reversal symmetry-breaking superconductivity in
  $\mathrm{Sr}_2\mathrm{RuO}_4$},\ }\href {https://doi.org/10.1038/29038}
  {\bibfield  {journal} {\bibinfo  {journal} {Nature}\ }\textbf {\bibinfo
  {volume} {394}},\ \bibinfo {pages} {558} (\bibinfo {year}
  {1998})}\BibitemShut {NoStop}%
\bibitem [{\citenamefont {Ishida}\ \emph {et~al.}(1998)\citenamefont {Ishida},
  \citenamefont {Mukuda}, \citenamefont {Kitaoka}, \citenamefont {Asayama},
  \citenamefont {Mao}, \citenamefont {Mori},\ and\ \citenamefont
  {Maeno}}]{ishida_knight_1998}%
  \BibitemOpen
  \bibfield  {author} {\bibinfo {author} {\bibfnamefont {K.}~\bibnamefont
  {Ishida}}, \bibinfo {author} {\bibfnamefont {H.}~\bibnamefont {Mukuda}},
  \bibinfo {author} {\bibfnamefont {Y.}~\bibnamefont {Kitaoka}}, \bibinfo
  {author} {\bibfnamefont {K.}~\bibnamefont {Asayama}}, \bibinfo {author}
  {\bibfnamefont {Z.~Q.}\ \bibnamefont {Mao}}, \bibinfo {author} {\bibfnamefont
  {Y.}~\bibnamefont {Mori}},\ and\ \bibinfo {author} {\bibfnamefont
  {Y.}~\bibnamefont {Maeno}},\ }\bibfield  {title} {\bibinfo {title}
  {Spin-triplet superconductivity in {${\mathrm{Sr}}_{2}{\mathrm{RuO}}_{4}$}
  identified by ${}^{17}${O} knight shift},\ }\href
  {https://doi.org/10.1038/25315} {\bibfield  {journal} {\bibinfo  {journal}
  {Nature}\ }\textbf {\bibinfo {volume} {396}},\ \bibinfo {pages} {658}
  (\bibinfo {year} {1998})}\BibitemShut {NoStop}%
\bibitem [{\citenamefont {Duffy}\ \emph {et~al.}(2000)\citenamefont {Duffy},
  \citenamefont {Hayden}, \citenamefont {Maeno}, \citenamefont {Mao},
  \citenamefont {Kulda},\ and\ \citenamefont
  {McIntyre}}]{duffy_polarized_2000}%
  \BibitemOpen
  \bibfield  {author} {\bibinfo {author} {\bibfnamefont {J.~A.}\ \bibnamefont
  {Duffy}}, \bibinfo {author} {\bibfnamefont {S.~M.}\ \bibnamefont {Hayden}},
  \bibinfo {author} {\bibfnamefont {Y.}~\bibnamefont {Maeno}}, \bibinfo
  {author} {\bibfnamefont {Z.}~\bibnamefont {Mao}}, \bibinfo {author}
  {\bibfnamefont {J.}~\bibnamefont {Kulda}},\ and\ \bibinfo {author}
  {\bibfnamefont {G.~J.}\ \bibnamefont {McIntyre}},\ }\bibfield  {title}
  {\bibinfo {title} {Polarized-neutron scattering study of the {Cooper}-pair
  moment in {${\mathrm{Sr}}_{2}{\mathrm{RuO}}_{4}$}},\ }\href
  {https://doi.org/10.1103/PhysRevLett.85.5412} {\bibfield  {journal} {\bibinfo
   {journal} {Phys. Rev. Lett.}\ }\textbf {\bibinfo {volume} {85}},\ \bibinfo
  {pages} {5412} (\bibinfo {year} {2000})}\BibitemShut {NoStop}%
\bibitem [{\citenamefont {Ishida}\ \emph {et~al.}(2001)\citenamefont {Ishida},
  \citenamefont {Mukuda}, \citenamefont {Kitaoka}, \citenamefont {Mao},
  \citenamefont {Fukazawa},\ and\ \citenamefont {Maeno}}]{ishida_spin_2001}%
  \BibitemOpen
  \bibfield  {author} {\bibinfo {author} {\bibfnamefont {K.}~\bibnamefont
  {Ishida}}, \bibinfo {author} {\bibfnamefont {H.}~\bibnamefont {Mukuda}},
  \bibinfo {author} {\bibfnamefont {Y.}~\bibnamefont {Kitaoka}}, \bibinfo
  {author} {\bibfnamefont {Z.~Q.}\ \bibnamefont {Mao}}, \bibinfo {author}
  {\bibfnamefont {H.}~\bibnamefont {Fukazawa}},\ and\ \bibinfo {author}
  {\bibfnamefont {Y.}~\bibnamefont {Maeno}},\ }\bibfield  {title} {\bibinfo
  {title} {Ru {NMR} probe of spin susceptibility in the superconducting state
  of {${\mathrm{Sr}}_{2}{\mathrm{RuO}}_{4}$}},\ }\href
  {https://doi.org/10.1103/PhysRevB.63.060507} {\bibfield  {journal} {\bibinfo
  {journal} {Phys. Rev. B}\ }\textbf {\bibinfo {volume} {63}},\ \bibinfo
  {pages} {060507} (\bibinfo {year} {2001})}\BibitemShut {NoStop}%
\bibitem [{\citenamefont {Pustogow}\ \emph {et~al.}(2019)\citenamefont
  {Pustogow}, \citenamefont {Luo}, \citenamefont {Chronister}, \citenamefont
  {Su}, \citenamefont {Sokolov}, \citenamefont {Jerzembeck}, \citenamefont
  {Mackenzie}, \citenamefont {Hicks}, \citenamefont {Kikugawa}, \citenamefont
  {Raghu}, \citenamefont {Bauer},\ and\ \citenamefont
  {Brown}}]{pustogow_constraints_2019}%
  \BibitemOpen
  \bibfield  {author} {\bibinfo {author} {\bibfnamefont {A.}~\bibnamefont
  {Pustogow}}, \bibinfo {author} {\bibfnamefont {Y.}~\bibnamefont {Luo}},
  \bibinfo {author} {\bibfnamefont {A.}~\bibnamefont {Chronister}}, \bibinfo
  {author} {\bibfnamefont {Y.-S.}\ \bibnamefont {Su}}, \bibinfo {author}
  {\bibfnamefont {D.~A.}\ \bibnamefont {Sokolov}}, \bibinfo {author}
  {\bibfnamefont {F.}~\bibnamefont {Jerzembeck}}, \bibinfo {author}
  {\bibfnamefont {A.~P.}\ \bibnamefont {Mackenzie}}, \bibinfo {author}
  {\bibfnamefont {C.~W.}\ \bibnamefont {Hicks}}, \bibinfo {author}
  {\bibfnamefont {N.}~\bibnamefont {Kikugawa}}, \bibinfo {author}
  {\bibfnamefont {S.}~\bibnamefont {Raghu}}, \bibinfo {author} {\bibfnamefont
  {E.~D.}\ \bibnamefont {Bauer}},\ and\ \bibinfo {author} {\bibfnamefont
  {S.~E.}\ \bibnamefont {Brown}},\ }\bibfield  {title} {\bibinfo {title}
  {Constraints on the superconducting order parameter in
  $\mathrm{Sr}_2\mathrm{RuO}_4$ from oxygen-17 nuclear magnetic resonance},\
  }\href {https://doi.org/10.1038/s41586-019-1596-2} {\bibfield  {journal}
  {\bibinfo  {journal} {Nature}\ }\textbf {\bibinfo {volume} {574}},\ \bibinfo
  {pages} {72} (\bibinfo {year} {2019})}\BibitemShut {NoStop}%
\bibitem [{\citenamefont {Ishida}\ \emph {et~al.}(2020)\citenamefont {Ishida},
  \citenamefont {Manago}, \citenamefont {Kinjo},\ and\ \citenamefont
  {Maeno}}]{ishida_reduction_2020}%
  \BibitemOpen
  \bibfield  {author} {\bibinfo {author} {\bibfnamefont {K.}~\bibnamefont
  {Ishida}}, \bibinfo {author} {\bibfnamefont {M.}~\bibnamefont {Manago}},
  \bibinfo {author} {\bibfnamefont {K.}~\bibnamefont {Kinjo}},\ and\ \bibinfo
  {author} {\bibfnamefont {Y.}~\bibnamefont {Maeno}},\ }\bibfield  {title}
  {\bibinfo {title} {Reduction of the ${}^{17}${O} {Knight} shift in the
  superconducting state and the heat-up effect by {NMR} pulses on
  $\mathrm{Sr}_2\mathrm{RuO}_4$},\ }\href
  {https://doi.org/10.7566/JPSJ.89.034712} {\bibfield  {journal} {\bibinfo
  {journal} {J. Phys. Soc. Jpn.}\ }\textbf {\bibinfo {volume} {89}},\ \bibinfo
  {pages} {034712} (\bibinfo {year} {2020})}\BibitemShut {NoStop}%
\bibitem [{\citenamefont {Tamai}\ \emph {et~al.}(2019)\citenamefont {Tamai},
  \citenamefont {Zingl}, \citenamefont {Rozbicki}, \citenamefont {Cappelli},
  \citenamefont {Ricc{\`o}}, \citenamefont {de~la Torre}, \citenamefont
  {McKeown~Walker}, \citenamefont {Bruno}, \citenamefont {King}, \citenamefont
  {Meevasana}, \citenamefont {Shi}, \citenamefont {Radovi{\'c}}, \citenamefont
  {Plumb}, \citenamefont {Gibbs}, \citenamefont {Mackenzie}, \citenamefont
  {Berthod}, \citenamefont {Strand}, \citenamefont {Kim}, \citenamefont
  {Georges},\ and\ \citenamefont {Baumberger}}]{tamai_high-resolution_2019}%
  \BibitemOpen
  \bibfield  {author} {\bibinfo {author} {\bibfnamefont {A.}~\bibnamefont
  {Tamai}}, \bibinfo {author} {\bibfnamefont {M.}~\bibnamefont {Zingl}},
  \bibinfo {author} {\bibfnamefont {E.}~\bibnamefont {Rozbicki}}, \bibinfo
  {author} {\bibfnamefont {E.}~\bibnamefont {Cappelli}}, \bibinfo {author}
  {\bibfnamefont {S.}~\bibnamefont {Ricc{\`o}}}, \bibinfo {author}
  {\bibfnamefont {A.}~\bibnamefont {de~la Torre}}, \bibinfo {author}
  {\bibfnamefont {S.}~\bibnamefont {McKeown~Walker}}, \bibinfo {author}
  {\bibfnamefont {F.~Y.}\ \bibnamefont {Bruno}}, \bibinfo {author}
  {\bibfnamefont {P.~D.~C.}\ \bibnamefont {King}}, \bibinfo {author}
  {\bibfnamefont {W.}~\bibnamefont {Meevasana}}, \bibinfo {author}
  {\bibfnamefont {M.}~\bibnamefont {Shi}}, \bibinfo {author} {\bibfnamefont
  {M.}~\bibnamefont {Radovi{\'c}}}, \bibinfo {author} {\bibfnamefont {N.~C.}\
  \bibnamefont {Plumb}}, \bibinfo {author} {\bibfnamefont {A.~S.}\ \bibnamefont
  {Gibbs}}, \bibinfo {author} {\bibfnamefont {A.~P.}\ \bibnamefont
  {Mackenzie}}, \bibinfo {author} {\bibfnamefont {C.}~\bibnamefont {Berthod}},
  \bibinfo {author} {\bibfnamefont {H.~U.~R.}\ \bibnamefont {Strand}}, \bibinfo
  {author} {\bibfnamefont {M.}~\bibnamefont {Kim}}, \bibinfo {author}
  {\bibfnamefont {A.}~\bibnamefont {Georges}},\ and\ \bibinfo {author}
  {\bibfnamefont {F.}~\bibnamefont {Baumberger}},\ }\bibfield  {title}
  {\bibinfo {title} {High-resolution photoemission on
  $\mathrm{Sr}_2\mathrm{RuO}_4$ reveals correlation-enhanced effective
  spin-orbit coupling and dominantly local self-energies},\ }\href
  {https://doi.org/10.1103/PhysRevX.9.021048} {\bibfield  {journal} {\bibinfo
  {journal} {Phys. Rev. X}\ }\textbf {\bibinfo {volume} {9}},\ \bibinfo {pages}
  {021048} (\bibinfo {year} {2019})}\BibitemShut {NoStop}%
\bibitem [{\citenamefont {Rozbicki}\ \emph {et~al.}(2011)\citenamefont
  {Rozbicki}, \citenamefont {Annett}, \citenamefont {Souquet},\ and\
  \citenamefont {Mackenzie}}]{rozbicki_spinorbit_2011}%
  \BibitemOpen
  \bibfield  {author} {\bibinfo {author} {\bibfnamefont {E.~J.}\ \bibnamefont
  {Rozbicki}}, \bibinfo {author} {\bibfnamefont {J.~F.}\ \bibnamefont
  {Annett}}, \bibinfo {author} {\bibfnamefont {J.-R.}\ \bibnamefont
  {Souquet}},\ and\ \bibinfo {author} {\bibfnamefont {A.~P.}\ \bibnamefont
  {Mackenzie}},\ }\bibfield  {title} {\bibinfo {title} {Spin{\textendash}orbit
  coupling and $k$-dependent {Zeeman} splitting in strontium ruthenate},\
  }\href {https://doi.org/10.1088/0953-8984/23/9/094201} {\bibfield  {journal}
  {\bibinfo  {journal} {J. Phys.: Condens. Matter}\ }\textbf {\bibinfo {volume}
  {23}},\ \bibinfo {pages} {094201} (\bibinfo {year} {2011})}\BibitemShut
  {NoStop}%
\bibitem [{\citenamefont {Haverkort}\ \emph {et~al.}(2008)\citenamefont
  {Haverkort}, \citenamefont {Elfimov}, \citenamefont {Tjeng}, \citenamefont
  {Sawatzky},\ and\ \citenamefont {Damascelli}}]{haverkort_strong_2008}%
  \BibitemOpen
  \bibfield  {author} {\bibinfo {author} {\bibfnamefont {M.~W.}\ \bibnamefont
  {Haverkort}}, \bibinfo {author} {\bibfnamefont {I.~S.}\ \bibnamefont
  {Elfimov}}, \bibinfo {author} {\bibfnamefont {L.~H.}\ \bibnamefont {Tjeng}},
  \bibinfo {author} {\bibfnamefont {G.~A.}\ \bibnamefont {Sawatzky}},\ and\
  \bibinfo {author} {\bibfnamefont {A.}~\bibnamefont {Damascelli}},\ }\bibfield
   {title} {\bibinfo {title} {Strong spin-orbit coupling effects on the {Fermi}
  surface of $\mathrm{Sr}_2\mathrm{RuO}_4$ and $\mathrm{Sr}_2\mathrm{RhO}_4$},\
  }\href {https://doi.org/10.1103/PhysRevLett.101.026406} {\bibfield  {journal}
  {\bibinfo  {journal} {Phys. Rev. Lett.}\ }\textbf {\bibinfo {volume} {101}},\
  \bibinfo {pages} {026406} (\bibinfo {year} {2008})}\BibitemShut {NoStop}%
\bibitem [{\citenamefont {Black-Schaffer}\ and\ \citenamefont
  {Doniach}(2007)}]{BlackSchaffer_graphite_2007}%
  \BibitemOpen
  \bibfield  {author} {\bibinfo {author} {\bibfnamefont {A.~M.}\ \bibnamefont
  {Black-Schaffer}}\ and\ \bibinfo {author} {\bibfnamefont {S.}~\bibnamefont
  {Doniach}},\ }\bibfield  {title} {\bibinfo {title} {Resonating valence bonds
  and mean-field $d$-wave superconductivity in graphite},\ }\href
  {https://doi.org/10.1103/PhysRevB.75.134512} {\bibfield  {journal} {\bibinfo
  {journal} {Phys. Rev. B}\ }\textbf {\bibinfo {volume} {75}},\ \bibinfo
  {pages} {134512} (\bibinfo {year} {2007})}\BibitemShut {NoStop}%
\bibitem [{\citenamefont {Wu}\ \emph {et~al.}(2013)\citenamefont {Wu},
  \citenamefont {Scherer}, \citenamefont {Honerkamp},\ and\ \citenamefont
  {Le~Hur}}]{Wu_2013}%
  \BibitemOpen
  \bibfield  {author} {\bibinfo {author} {\bibfnamefont {W.}~\bibnamefont
  {Wu}}, \bibinfo {author} {\bibfnamefont {M.~M.}\ \bibnamefont {Scherer}},
  \bibinfo {author} {\bibfnamefont {C.}~\bibnamefont {Honerkamp}},\ and\
  \bibinfo {author} {\bibfnamefont {K.}~\bibnamefont {Le~Hur}},\ }\bibfield
  {title} {\bibinfo {title} {Correlated {Dirac} particles and superconductivity
  on the honeycomb lattice},\ }\href
  {https://doi.org/10.1103/PhysRevB.87.094521} {\bibfield  {journal} {\bibinfo
  {journal} {Phys. Rev. B}\ }\textbf {\bibinfo {volume} {87}},\ \bibinfo
  {pages} {094521} (\bibinfo {year} {2013})}\BibitemShut {NoStop}%
\bibitem [{\citenamefont {Kane}\ and\ \citenamefont
  {Mele}(2005)}]{Kane_Mele_2005}%
  \BibitemOpen
  \bibfield  {author} {\bibinfo {author} {\bibfnamefont {C.~L.}\ \bibnamefont
  {Kane}}\ and\ \bibinfo {author} {\bibfnamefont {E.~J.}\ \bibnamefont
  {Mele}},\ }\bibfield  {title} {\bibinfo {title} {Quantum spin {Hall} effect
  in graphene},\ }\href {https://doi.org/10.1103/PhysRevLett.95.226801}
  {\bibfield  {journal} {\bibinfo  {journal} {Phys. Rev. Lett.}\ }\textbf
  {\bibinfo {volume} {95}},\ \bibinfo {pages} {226801} (\bibinfo {year}
  {2005})}\BibitemShut {NoStop}%
\bibitem [{\citenamefont {Suh}\ \emph {et~al.}(2020)\citenamefont {Suh},
  \citenamefont {Menke}, \citenamefont {Brydon}, \citenamefont {Timm},
  \citenamefont {Ramires},\ and\ \citenamefont
  {Agterberg}}]{SuhMenke_SRO_2020}%
  \BibitemOpen
  \bibfield  {author} {\bibinfo {author} {\bibfnamefont {H.~G.}\ \bibnamefont
  {Suh}}, \bibinfo {author} {\bibfnamefont {H.}~\bibnamefont {Menke}}, \bibinfo
  {author} {\bibfnamefont {P.~M.~R.}\ \bibnamefont {Brydon}}, \bibinfo {author}
  {\bibfnamefont {C.}~\bibnamefont {Timm}}, \bibinfo {author} {\bibfnamefont
  {A.}~\bibnamefont {Ramires}},\ and\ \bibinfo {author} {\bibfnamefont {D.~F.}\
  \bibnamefont {Agterberg}},\ }\bibfield  {title} {\bibinfo {title}
  {Stabilizing even-parity chiral superconductivity in
  $\mathrm{Sr}_{2}\mathrm{RuO}_{4}$},\ }\href
  {https://doi.org/10.1103/PhysRevResearch.2.032023} {\bibfield  {journal}
  {\bibinfo  {journal} {Phys. Rev. Research}\ }\textbf {\bibinfo {volume}
  {2}},\ \bibinfo {pages} {032023} (\bibinfo {year} {2020})}\BibitemShut
  {NoStop}%
\bibitem [{\citenamefont {Hillier}\ \emph {et~al.}(2009)\citenamefont
  {Hillier}, \citenamefont {Quintanilla},\ and\ \citenamefont
  {Cywinski}}]{hillier_evidence_2009}%
  \BibitemOpen
  \bibfield  {author} {\bibinfo {author} {\bibfnamefont {A.~D.}\ \bibnamefont
  {Hillier}}, \bibinfo {author} {\bibfnamefont {J.}~\bibnamefont
  {Quintanilla}},\ and\ \bibinfo {author} {\bibfnamefont {R.}~\bibnamefont
  {Cywinski}},\ }\bibfield  {title} {\bibinfo {title} {Evidence for
  time-reversal symmetry breaking in the noncentrosymmetric superconductor
  {${\mathrm{LaNiC}}_{2}$}},\ }\href
  {https://doi.org/10.1103/PhysRevLett.102.117007} {\bibfield  {journal}
  {\bibinfo  {journal} {Phys. Rev. Lett.}\ }\textbf {\bibinfo {volume} {102}},\
  \bibinfo {pages} {117007} (\bibinfo {year} {2009})}\BibitemShut {NoStop}%
\end{thebibliography}%

\end{document}